\documentclass{aa}     
\usepackage{psfig}

\newcommand{\mum}{\,$\rm\mu m$}            
\newcommand{\z}{\phantom{0}}               

\newcommand{\struth}{\rule[-1ex]{0mm}{3.5ex}}       
\newcommand{\strutb}{\rule[-1.5ex]{0mm}{1.5ex}}       
\newcommand{\strutt}{\rule{0mm}{2.8ex}}             

\begin{document}

\font\myfontsmall=cmr17 at 8.0pt

   \thesaurus{04.19.1, 04.03.1, 13.25.3
             }

   \title{The ROSAT All-Sky Survey Bright Source Catalogue}

   \author{ W. Voges \and B. Aschenbach \and  
Th. Boller \and H. Br\"auninger \and U. Briel \and W. Burkert \and 
K. Dennerl \and J. Englhauser \and R. Gruber \and F. Haberl \and 
G. Hartner \and G. Hasinger\thanks{\emph{Present address:} 
Astrophysikalisches Institut Potsdam, 
An der Sternwarte 16, 14482 Potsdam, Germany}
\and M. K\"urster\thanks{\emph{Present address:} ESO--La Silla, 
Alonso de Cordova 3107, Vitacura, Casilla 19001, Santiago 19, Chile}
\and E. Pfeffermann \and 
W. Pietsch \and P. Predehl \and C. Rosso 
\and J.H.M.M. Schmitt\thanks{\emph{Present 
address:} Hamburger Sternwarte, Gojenbergsweg 112, 21029 
Hamburg, Germany} \and J. Tr\"umper \and  H.U. Zimmermann}    

   \date{Received 1 March 1999 /  Accepted 25 June 1999}

   \institute{Max-Planck-Institut f\"ur extraterrestrische Physik,
                   Postfach 1603, 85740 Garching, Germany}

   \offprints{W. Voges (wvoges@mpe.mpg.de)}

   \date{Received 1 March 1999 /  Accepted 25 June 1999}
   \maketitle
   \begin{abstract}
We present the ROSAT All-Sky Survey Bright Source Catalogue (RASS-BSC, revision
1RXS) derived from the all-sky survey performed during the first half year
(1990/91) of the ROSAT mission. 18,811 sources are catalogued (i) down to a
limiting ROSAT PSPC count-rate of 0.05 cts/s in the 0.1$-$2.4 keV energy band,
(ii)  with a detection likelihood of at least 15 and (iii) at least 15 source
counts. The 18,811 sources underwent both an automatic validation and an
interactive visual verification process in which for 94\% of the sources the
results of the standard processing were confirmed. The remaining 6\% have been
analyzed using interactive methods and these sources have been flagged. Flags
are given for (i) nearby sources; (ii) sources with positional errors; (iii)
extended sources; (iv) sources showing complex emission structures; and (v)
sources which are missed by the standard analysis software. Broad band
(0.1$-$2.4 keV) images are available for sources flagged by (ii), (iii) and
(iv). For each source the ROSAT name, position in equatorial coordinates,
positional error, source count-rate and error, background count-rate, exposure
time, two hardness-ratios and errors, extent and likelihood of extent,
likelihood of detection, and the source extraction radius are provided. At a
brightness limit of 0.1 cts/s (8,547 sources) the catalogue represents a sky
coverage of 92\%. The RASS-BSC, the table of possible identification
candidates, and the broad band images are available in electronic form (Voges
et al. 1996a) via http://wave.xray.mpe.mpg.de/rosat/catalogues/rass-bsc .
\footnote{ The RASS-BSC and the identification table are also available in
electronic form at the CDS via anonymous ftp to cdsarc.u-strasbg.fr
(130.79.128.5) or via http://cdsweb.u-strasbg.fr/Abstract.html}

\vspace{-.25cm}


\end{abstract}

\section{Introduction}

\begin{table*}
\begin{center}
\begin{tabular}{lccrl}
 \hline
Satellites & date & energy      & number of & References  \\
           &      & range (keV) &  sources  &   \\
  \hline
 UHURU                  & 1970-73 &  2-6    & 339  & Forman et al. (1978)  \\
 OSO-7                  & 1971-73 & 1-60    & 184  & Markert et al. (1979)  \\
 ARIEL-5 ($|$ b$^{\rm II}$ $|$ $>$10$\degr$) & 1974-80 & 2-18  & 142  & McHardy et al. (1981)  \\
 ARIEL-5 ($|$ b$^{\rm II}$ $|$ $<$10$\degr$) & 1974-80 & 2-10  & 109  & Warwick et al. (1981)  \\
 HEAO-1/A1              & 1977-79 & 1-20    & 842  & Wood et al. (1984)  \\
 HEAO-1/A2              & 1977-79 & 0.2-2.8 & 114  & Nugent et al. (1883) \\
 HEAO-1/A4              & 1977-79 & 13-180  &  40  & Levine et al. (1984)  \\
\hline
\end{tabular}
\caption{X-ray astronomy missions and all-sky surveys before the launch of ROSAT.}
\end{center}
\end{table*}

Sky surveys play  a major role in observational astronomy, in particular in the
era of multi-wavelength observations. Before the launch of the ROSAT satellite
several all-sky X-ray catalogues existed based on collimated counter surveys
(see Table 1). One of the main scientific objectives of ROSAT was to conduct
the first all-sky survey in X-rays with an imaging telescope leading to a
major increase in sensitivity and source location accuracy (Tr\"umper 1983).
Actually, the ROSAT mirror system (Aschenbach 1988) and the Position Sensitive
Proportional Counter (PSPC) (Pfeffermann et al. 1988) used for the survey were
primarily optimized for detecting point sources in the all-sky survey.
However,  the wide angle and fast optics  as well as the low detector
background of the PSPC provided excellent conditions for studying extended
sources like supernova remnants, clusters of galaxies, and the diffuse X-ray
background. In this context the ``unlimited field of view" of the all-sky
survey was of great advantage.
 
The ROSAT All-Sky Survey (RASS) was conducted in 1990/91, just after the two
month switch-on and performance verification phase. The first processing of
the ROSAT All-Sky Survey took place in 1991$-$1993 resulting in about 50,000
sources. This source list had not been published, because during the analysis
a large number of minor deficiencies and possibilities for improvement were
discovered. Nevertheless, the data have been extensively used by the
scientific groups at MPE and their collaborators for many scientific projects.

Based on the experience with the first RASS processing a second analysis was
performed in 1994$-$1995, resulting in 145,060  sources (detection likelihood
$\ge$ 7). The present publication comprises only the brightest 18,811 of this
sample. These data represent by far the most complete and sensitive X-ray sky
survey ever published. It is a factor of 20 more sensitive than any previous
all-sky survey in X-rays and contains about a factor of 4 more sources than
all other X-ray catalogues, which sample only a few percent of the sky.
 
In this paper we present the outcome of the RASS in terms of ``point sources".
For 98.2\% of the 18,811 sources the source extent radius is less than 5
arcmin, and for 99.6\% the extent is smaller than 10 arcmin. 0.4\% of the
sources exhibit a larger source extent and show complex emission patterns (see
Section 3.2) rather than a point-like spatial distribution. These sources have
been included for completeness as they fulfill the selection criteria of the
RASS-BSC (hereafter RBSC). Diffuse sky maps with angular resolution of 40
arcmin and diffuse sky maps in six colours of 12 arcmin resolution have been
published elsewhere (Snowden et al. 1995, 1997).
 
In Sect. 2 we summarize the basic properties of the ROSAT All-Sky Survey and
of the Standard Analysis Software System (hereafter SASS). The selection
strategy for including sources into the RBSC, the screening process, and the
content of the RBSC are presented in Sect. 3. The results from the correlation
of RBSC sources with various databases are given in Sect. 4. The electronic
access to the RBSC is described in Sect. 5.

\section{The ROSAT All-Sky Survey}

\subsection{Observation strategy and exposure map}
ROSAT has conducted the first All Sky Surveys in soft X-rays (0.1$-$2.4 keV;
100$-$5 \AA) and the extreme ultraviolet (0.025$-$0.2 keV; 500$-$60 \AA) bands
using imaging telescopes (Tr\"umper 1983, Aschenbach 1988, Wells et al. 1990,
Kent et al. 1990). The satellite was launched on June 1, 1990 and saw first
light on June 16, 1990 (Tr\"umper et al. 1991). The following 6 week
calibration and verification phase already included a small fraction of the
sky survey (see Table 2). The main part of the survey began on July 30, 1990
and lasted until January 25, 1991. A strip of the sky which remained
uncovered, because of operational problems in January, was retrieved in
February and August 1991 (see Table 2). The data obtained until 1991 form the
basis of the present analysis. The total survey exposure time amounts to 1.031
$\cdot$ 10$^{7}$s or 119.36 days.

The basic survey strategy of ROSAT was to scan the sky in great circles whose
planes were oriented roughly perpendicular to the solar direction. This
resulted in an exposure time varying between about 400 s and 40,000 s at the
ecliptic equator and poles respectively. During the passages through the
auroral zones and the South Atlantic Anomaly the PSPC had to be switched off,
leading to a decrease of exposure over parts of the sky (see Fig. 1). The sky
coverage as a function of the exposure time is displayed in Fig. 2. For
exposure times larger than 50 seconds the sky coverage is 99.7\% for the
observations until 1991.

\begin{table}
\begin{center}
\begin{tabular}{c@{ }c@{ }c@{ - }c@{ }c@{ }ccc}
\hline
            \multicolumn{6}{c}{Dates}      & ROSAT\, days & Detector \\
            \hline
            1990&Jul&11 & 1990&Jul&16      & \z41--\z45  & PSPC-C \\
            1990&Jul&30 & 1991&Jan&25      & \z60--242   & PSPC-C \\
            1991&Feb&16 & 1991&Feb&18      & 263--264    & PSPC-B \\
            1991&Aug&03 & 1991&Aug&13      & 435--444    & PSPC-B \\
             \hline
\end{tabular}
\end{center}
      \caption{ROSAT All-Sky Survey observation intervals.}
\end{table}

\subsection{SASS processing}
The first analysis of the all-sky survey data was performed for strips of
$2\degr\times360\degr$ containing the data taken during two days. These strips
were analysed using various source detection algorithms, comprising two
sliding window techniques (differing in how the background was determined) and
a maximum-likelihood method. A list of  X-ray sources (RASS-I) was produced
which included information about sky position and source properties, such as
count-rate, hardness-ratios, extent, and source detection likelihoods. The
main aim of this analysis was to supply almost immediate information about the
X-ray sources and to allow a fast quality check of the survey performance.

The RBSC presented in this paper is based on the so-called RASS-II processing
which is described below in Sect. 2.2.1--2.2.3.

\begin{figure}
  \centerline{\psfig{file=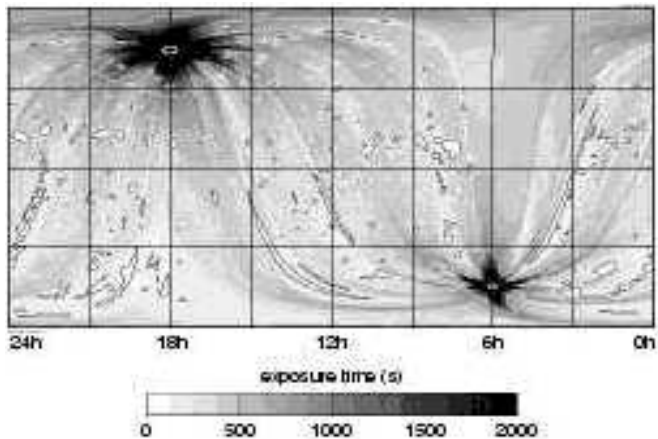,width=8.7cm,clip=}}
  \caption[Nichts]{\label{xm}
      Exposure map after the second processing of the ROSAT All-Sky Survey data,
      displayed in equatorial coordinates / rectangular projection.
      The spatial resolution of the map is 900 arcsec.
      Contours are drawn at 100\,s, 1000\,s, and 10\,000\,s, respectively.
      Due to the survey scan law the ecliptic poles 
      (at R.A.= 6 h and R.A. = 18 h) received the highest exposure.
      The southern hemisphere on average received less exposure 
      due to missed data during South Atlantic Anomaly passages.
      At some locations in the sky (seen as white spots) there were not 
      enough guide stars for
      the automatic measuring and control system of the satellite.
     }
\end{figure}

\begin{figure}
  \centerline{\psfig{file=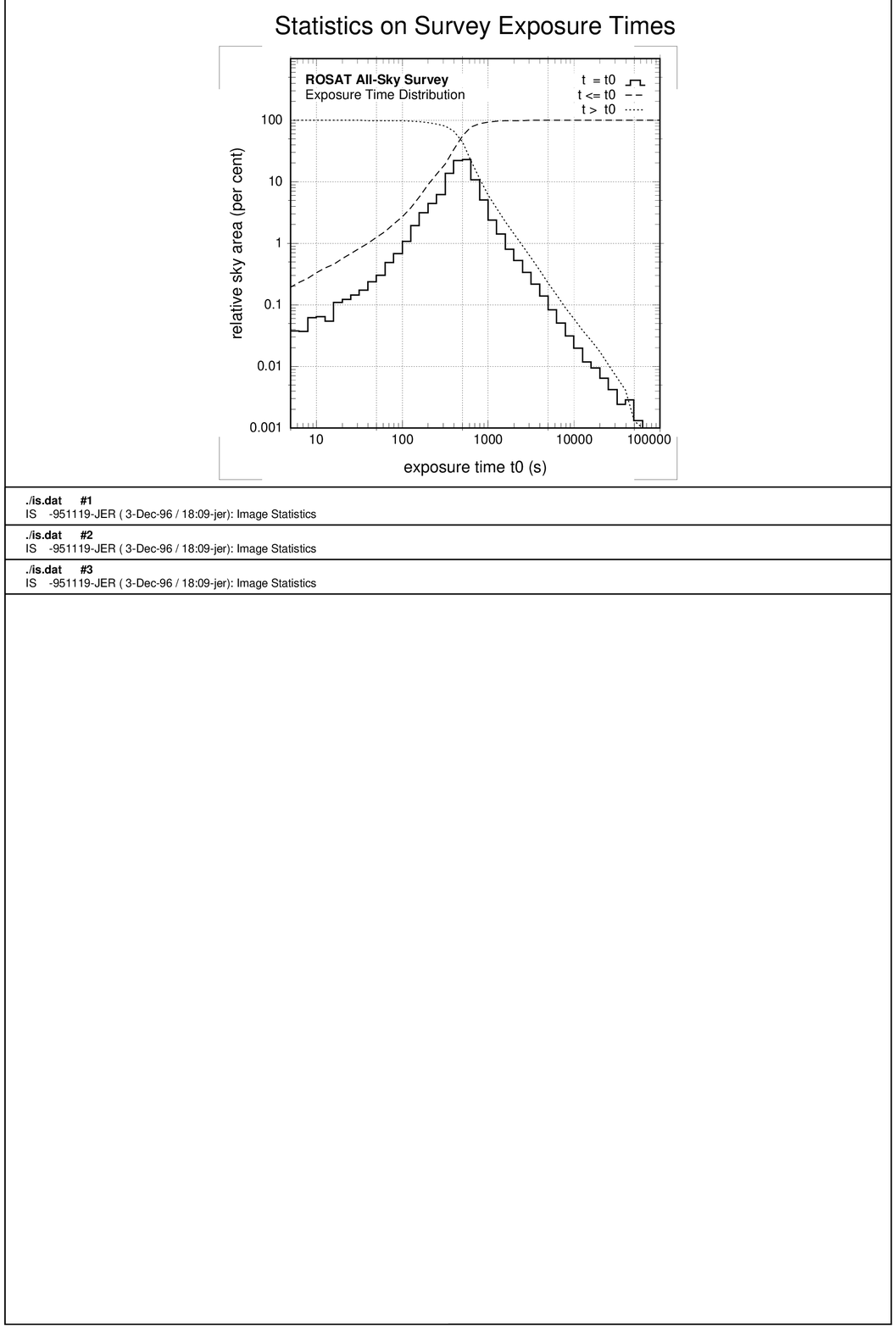,width=8.7cm,clip=}}
  \caption[Nichts]{\label{is}
      Differential and cumulative exposure time distribution for the
      second processing of the ROSAT All-Sky Survey.
      The time bins are equidistantly spaced logarithmically.
      The histogram shows the fraction of the sky covered 
      with exposure time $\rm t0$.
      The dashed and dotted lines show the fraction of the sky with 
      exposure time less than $\rm t0$ or greater than $\rm t0$, respectively.
      The statistics were derived from the all-sky exposure map shown
      in Fig.~\ref{xm} (resolution $15'$).
      For most of the sky ($97\%$) the exposure time was more than $\rm 100\ s$.
      The regions around the ecliptic poles received the highest
      exposure: 27 square degrees received more 
      than $\rm 10,000\ s$ and
      135 square degrees received less than $\rm 5\ s$.
     }
\end{figure}

\subsubsection{Advantages of the RASS-II processing}
The main differences between the RASS-II data processing and the RASS-I
processing are as follows: (i) the photons were not collected in strips but
were merged in 1,378 sky-fields of size $6.4\degr\times6.4\degr$, so that full
advantage was taken of the increasing exposure towards the ecliptic poles;
(ii) neighbouring fields overlapped by at least 0.23 degrees, to ensure
detection of sources at the field boundaries, which posed a problem in the
first processing; (iii) a new aspect solution reduced the number of sources
with erroneous position and morphology; (iv) the calculation of the
spline-fitted background map was improved, resulting in better determined
count-rates; (v) the candidate list for the maximum-likelihood analysis (see
Sect. 2.2.2) was enlarged by lowering the threshold values for the two
preceding sliding window source detection algorithms, and by changing the
acceptance criteria to allow very soft and very hard sources to be included;
and (vi) photons obtained with poor aspect solutions were no longer accepted.

\subsubsection{Source detection algorithm}
The source detection algorithms of the SASS processing can be divided
into 6 different steps:

\parindent=0.0cm
\vskip 0.2cm
1. The local-detect method\\
The local-detect algorithm is based on a sliding window technique. It was
already successfully used for the analysis of EINSTEIN data and has been
modified for ROSAT. A window of $3 \times 3$ pixels is moved across the binned
photon images. These images are produced by binning the data into 512 by 512
pixel images with a pixel size of 45 arcsec for three energy bands (broad:
Pulse Height Amplitude (PHA) channels 11--235 (0.1$-$2.4 keV), soft: channels
11--41 (0.1$-$0.4 keV), hard: channels 52--201 (0.5$-$2.0 keV)). The contents
of the pixels inside the detection cell are added and compared with the local
background taken from the 16 pixels surrounding the $\rm 3 \times 3$ pixel
detection window. To detect also extended sources the size of the detection
cell is increased systematically by keeping the ratio of the areas of
background and source windows fixed at 16/9.

\vskip 0.2cm
2. The background map\\
Using the source list produced in three energy bands by the local-detect
method, circular regions are cut out around each source position. The radius
of the circle is dependent on the detection cell size. The resulting
``swiss-cheese''-images are fitted by a two-dimensional spline function to
fill the holes and to generate three energy dependent background maps. In Fig.
3 we show the background maps in the soft and hard energy bands in galactic
coordinates.

\vskip 0.2cm
3. The map-detect algorithm\\
The map-detect algorithm produces a second source list by repeating the
sliding window search, using a $\rm 3 \times 3$ pixel window and the spline
fit to the background. Again sources are searched for in three energy bands
and with varying cell size.

\vskip 0.2cm
4. Merging of source lists\\
The source lists from the local- and map-detection algorithms are 
merged and are further used as input lists to steps 5 and 6.

\vskip 0.2cm
5. Determination of source extraction radius\\
A  preliminary extent of the source counts is derived from the radial
distribution of counts in annuli centered on the source position. This extent,
with a minimum being fixed at 300 arcsec, is baselined as an extraction radius
for the selection of the photons used in the subsequent maximum-likelihood
detection algorithm.

\vskip 0.2cm
6. The maximum-likelihood method\\
The merged source lists are used as input to the maximum-likelihood method
(Cruddace et al. 1987). In contrast to the previously described detection
algorithms this method takes into account the position of each individual
photon. This allows a proper weighting of each photon with the instrument
point spread function, which is a strong function of off-axis angle. The
high-resolution photons in the center of the PSPC are weighted higher than the
off-axis photons. The maximum-likelihood method provides a source position and
existence likelihood in the broad band. With this position fixed, the
detection likelihood in each of the 4 energy bands A (PHA channels 11--41), B
(52--201), C (52--90), D (91--201) is calculated. Vignetting is taken into
account for each photon using an analytic fit, leading to a mean vignetting
factor for each source. Source extent and its likelihood are derived using
just the broad band data, assuming that the point response function and the
surface brightness are 2-D Gaussian functions, that they are independent of
photon energy, and that the background is uniform in annular rings concentric
around the optical axis.

For strong sources various techniques are applied in the SASS to quantify the
likelihood for time variability and to perform spectral fits. This information
is not included in the present version of the RBSC.

\subsubsection{Parameters derived from the RASS-II processing}
In the following section a few basic source parameters from the RASS-II 
processing
are described. A complete description  of the derived source parameters is
given at\\
{\it  http://wave.xray.mpe.mpg.de/rosat/documentation/ \\productguide}\\
and a description of the catalogue entries of the RBSC is available at\\
{\it  http://wave.xray.mpe.mpg.de/rosat/catalogues/rass-bsc}. \\ 
A source count-rate corrected for vignetting is given in the broad band. Two
hardness ratios HR1 and HR2 are calculated, which represent X-ray colours.
From the source counts in the band A and the band B HR1 is given by:
HR1=(B--A)/(B+A). HR2 is determined from the source counts in the bands C and
D by: HR2=(D--C)/(D+C). Since background subtraction is involved, the source
counts in some bands may be negative. These negative counts have been set to
zero, so that HR1 or HR2 becomes --1 or +1. Note that HR2 is a hardness ratio
constructed in the hard region, since bands C and D together contain the same
channels as the hard band B. Thus HR1 near --1 and HR2 near +1 is no
contradiction. Hardness ratio errors greater than 9.99 have been set to 9.99.

Each source has been assigned a 'priority' parameter, the leftmost 6
characters of which denote the detection history before the maximum-likelihood
algorithm was applied. These are: 1=M--broad, 2=L--broad, 3=M--hard,
4=L--hard, 5=M--soft, 6=L--soft. Here M and L stand for the map-detect
algorithm and the local-detect method, respectively. Broad, hard and soft
refer to the energy bands defined above. A flag is set to 0 for no detection
or 1 for detection.

The source extent ({\tt ext} in the RBSC) is defined as the excess above the
width of the point spread function given in arcsec. In addition, the
likelihood ({\tt extl}) for the source extent and the extraction radius in
arcsec for a source ({\tt extr}) used in the maximum-likelihood method are
provided.

\section{The ROSAT Bright Source Catalogue}

\subsection{Selection criteria}
The total number of sources found in the RASS II is 145,060 (detection
likelihood $\ge$ 7). From this database the RBSC  was selected according to
the following criteria: (i) the detection likelihood is $\geq$ 15; (ii) the
number of source photons is $\geq$ 15 and (iii) the source count-rate in the
(0.1$-$2.4 keV) energy band is $\geq$ 0.05 $\rm counts\ s^{-1}$, resulting in
23,394 sources. These sources underwent an intensive screening process, which
is described in the following section.

\begin{figure*}

\begin{center}
\mbox{
  \psfig{file=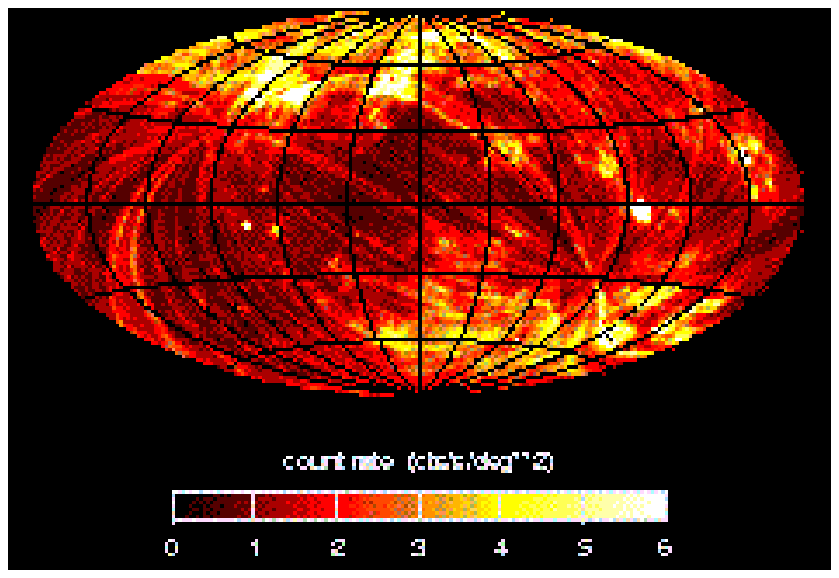,width=8.7cm,clip=}
  \psfig{file=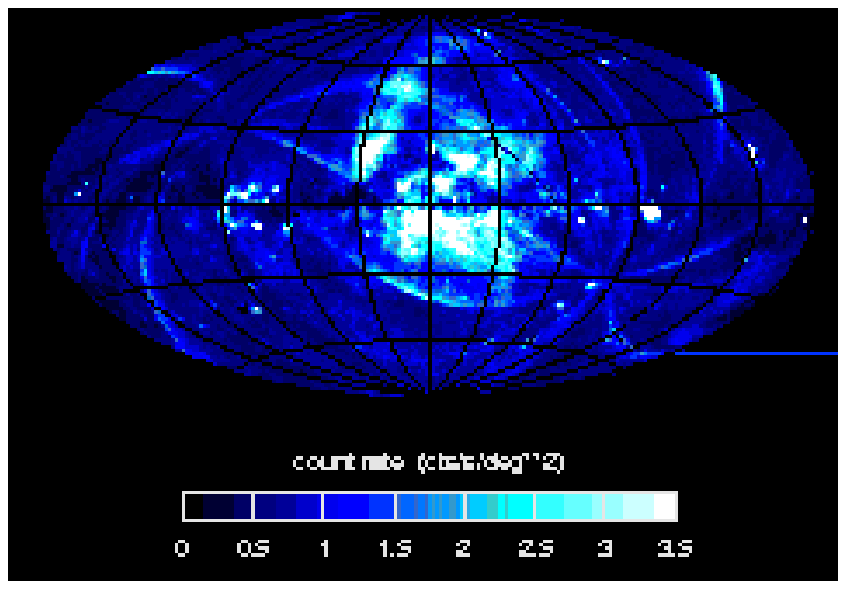,width=8.7cm,clip=}}
\end{center}
\vspace*{-3ex}
\caption{ Soft (left) and hard (right) background images in galactic 
          coordinates. The intensity is colour-coded.}

\begin{center}
\mbox{
\psfig{figure=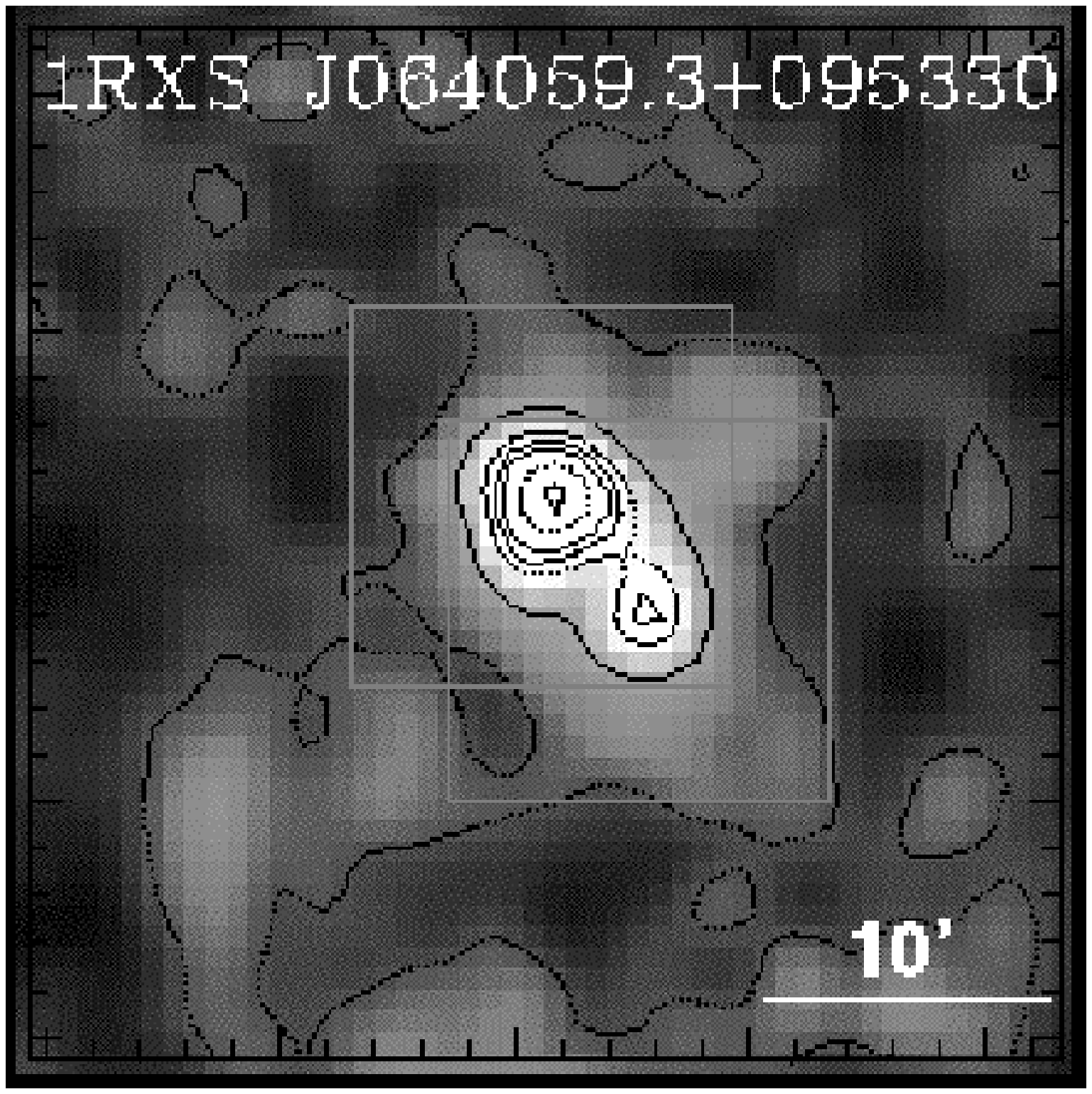,height=4.5cm,clip=}
\psfig{figure=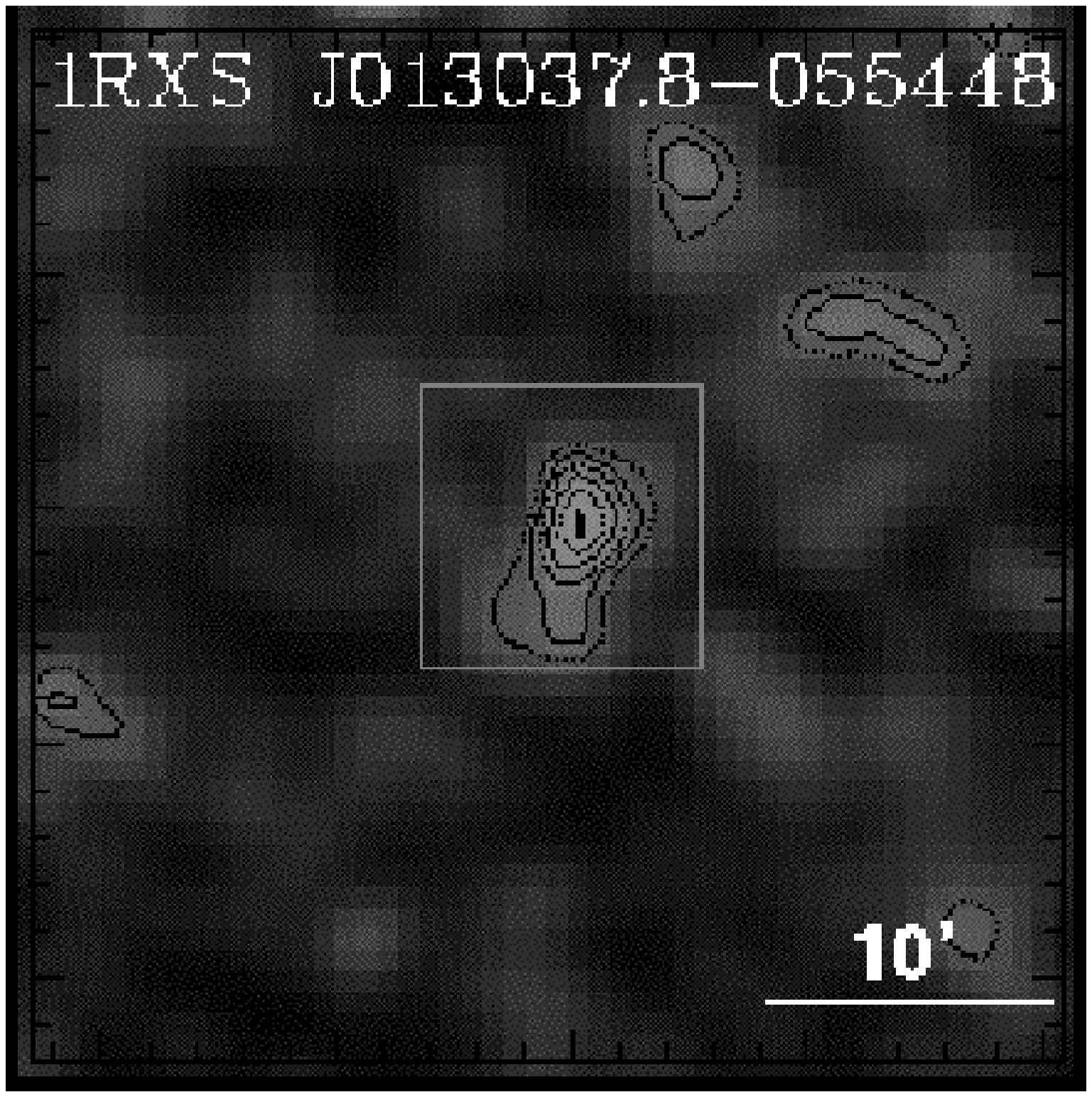,height=4.5cm,clip=}
\psfig{figure=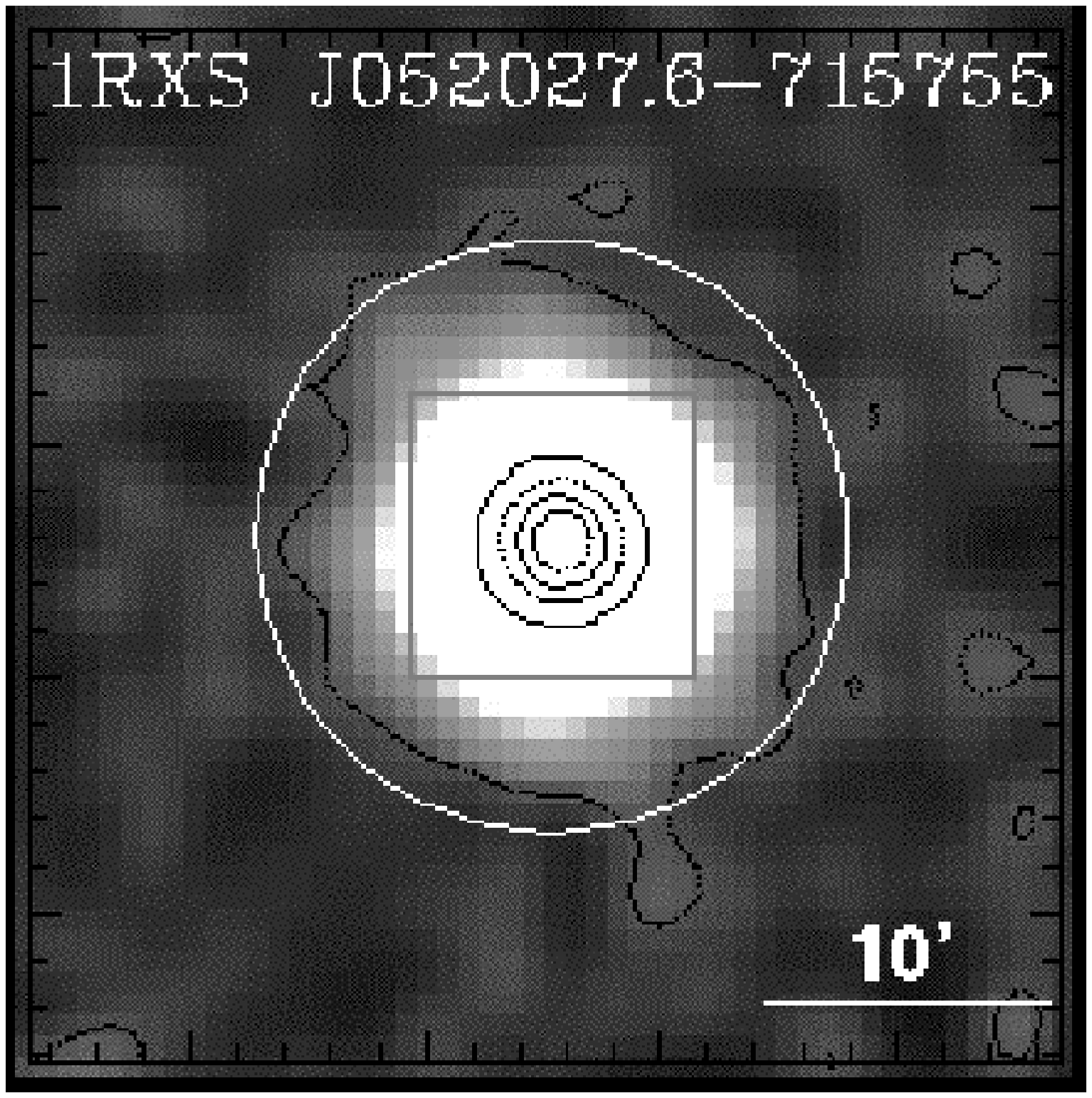,height=4.5cm,clip=}
\psfig{figure=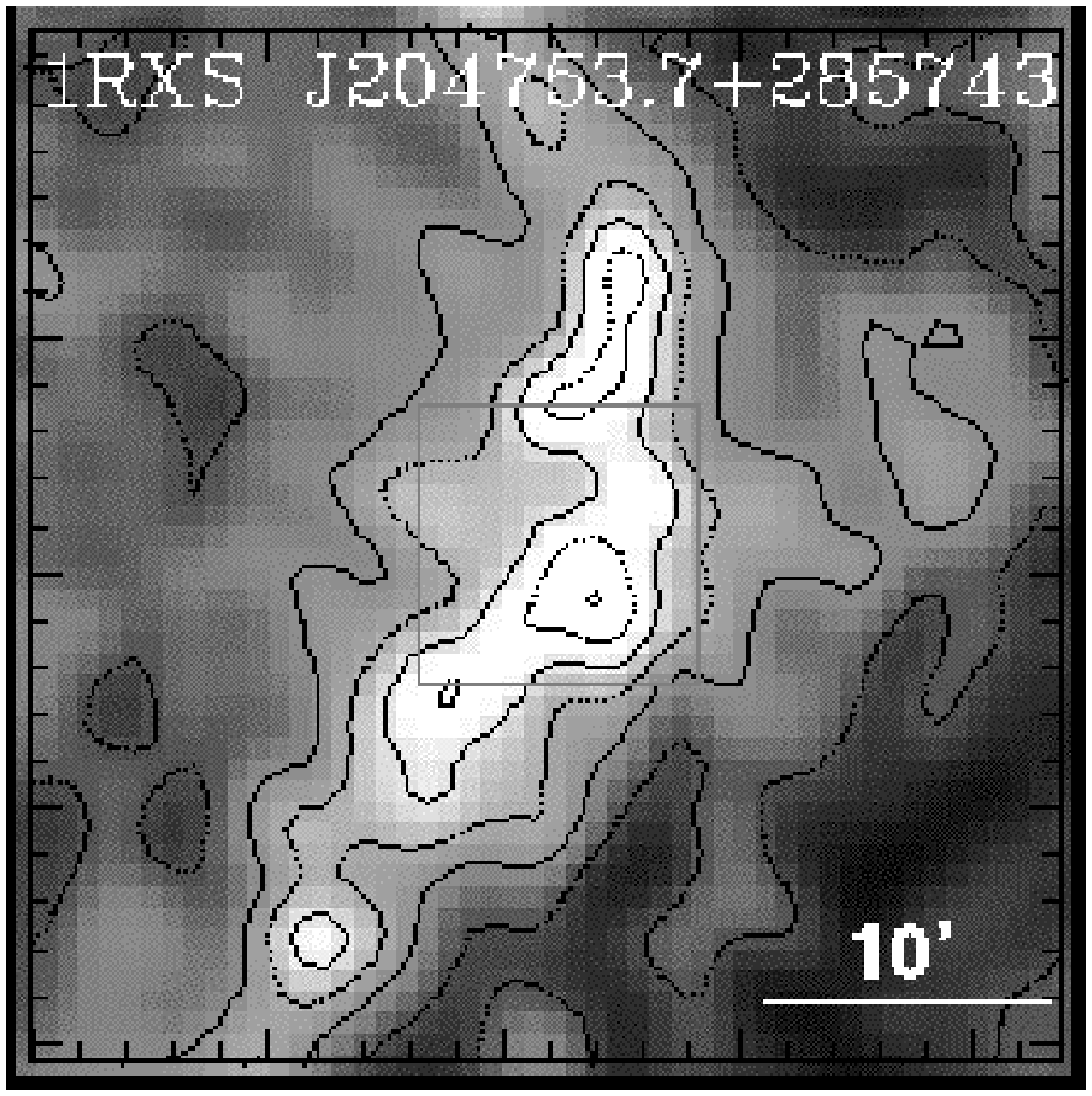,height=4.5cm,clip=}
}
\end{center}
\vspace*{-3ex}
\caption{ Examples for sources with warning flags. 
     X-ray contour lines are overlaid on the broad-band images.
     The SASS extraction radius is defined by the green rectangles and
     the new extraction radius, when determined, is shown by the white
     circle around the source position. From left to right:
     {\bf 1}: SASS flux determination affected by nearby sources;\ \
     {\bf 2}: Possible problem with position determination;
     {\bf 3}: Source extent larger than SASS extraction radius; \ \
     {\bf 4}: Complex diffuse emission pattern.
     }

\vspace*{2ex}
\centerline{\psfig{file=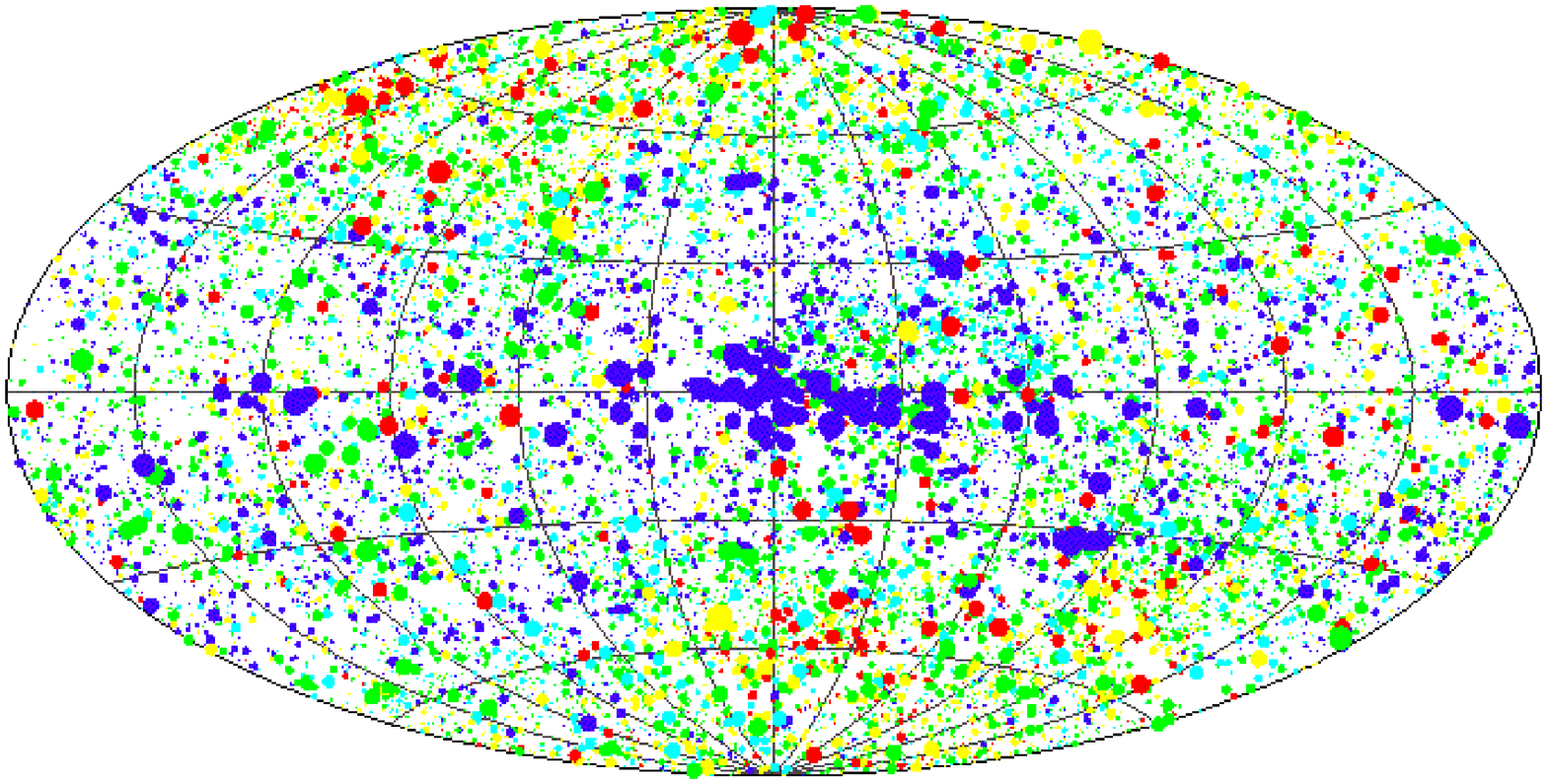,width=18.0cm,angle=-90,clip=}}
\caption[Nichts]{\label{aitoff}
     Aitoff projection of the distribution of all RBSC sources
     obtained in the ROSAT All-Sky Survey observations until August 13, 1991
     in galactic coordinates.
     The size of the symbols scales with the logarithm
     of the count-rate and the colours represent 5 intervals of the
     hardness ratio HR1:
     red ($\rm -1 \le HR1 < -0.6$);
     yellow ($\rm -0.6 \le HR1 < -0.2$);
     green ($\rm -0.2 \le HR1 < 0.2$);
     blue ($\rm 0.2 \le HR1 < 0.6$) and
     violet ($\rm 0.6 \le HR1 \le  1.0$).
     }

\end{figure*}

\subsection{Screening process}
In order to ensure a high quality of the catalogue, the images of all sources
which fulfill the above mentioned criteria were individually inspected.

An automatic as well as a visual screening procedure was applied to all 1,378
sky fields. The automatic procedure searched for sources with overlapping
extraction radii, as the count-rate determination can be affected in such
cases. These sources were colour-coded in the subsequent visual inspection
process.

The visual inspection process is based on ROSAT All-Sky Survey images in the
broad, soft and hard energy bands. The SASS position for each RBSC source as
well as the extraction radius were marked in the various images. This enables
the identification of regions on the sky where the detection algorithm had
split sources into multiple detections, and verifies that the source
extraction radius and the source position are correct. In addition, sources
which were missed by the detection algorithm can be found. During the visual
screening process source parameters from the SASS  could be checked
interactively, using software tools from the Extended Scientific Analysis
System (EXSAS, see Zimmermann et al. 1994). The reliability of the screening
process was verified in different ways. 100 of the 1,378 sky fields were used
as training sets for the screening process and were analysed by more than one
person to minimize the deviations in the flag setting. The results of the flag
setting were again visually inspected.

About 16\% of the 23,394 sources  were found to be spurious detections (mainly
in large extended emission regions like the Vela Supernova remnant and the
Cygnus Loop) which have been removed from the final source list. The remaining
18,811 sources are included in the present version of the RBSC. For 94\% of
these sources the SASS parameters were confirmed, and the remaining 6\% of the
sources have been flagged. The source flags applied to RBSC sources are
defined in the following paragraphs (see Fig. 4 for examples of flagged RBSC
sources).

{\tt (i) nearby flag}\\
This flag is set when the distance between two sources is less than the sum of
the individual extraction radii. The count-rate might be wrong in such cases.
The {\tt nearby flag} is given to both sources, when their count-rate ratio is
less than a factor of 5. If one source exhibits a count-rate which is at least
5 times higher than that of the weaker source, the nearby flag is given only
to the weaker source. In such cases, the count-rate of the brighter source is
not affected significantly. The nearby flag was given to 588 sources.

{\tt (ii) position error}\\
Whenever the source position is obviously not centered on the source
extraction cell the position flag is applied. We did not correct the positions
as there may be several different reasons for this problem (e.g. asymmetric
elongated emission patterns, multiple emission maxima in the detection cell).
The {\tt position error} flag was applied to 317 sources and a broad band
image provided for each one.

{\tt (iii) source extent larger than the source cell \\ 
size extraction radius.}\\
The standard analysis software fails in some cases to find the correct source
extent, with the result that the count-rates are usually underestimated. All
of these marked sources underwent a post-processing step to quantify whether
the source extent is indeed larger than the value found by the standard
analysis software. The number of sources marked with the {\tt source extent
flag} was 225. A broad band image is available for inspection and the new
extraction radius is indicated by a white circle.

{\tt (iv) complex emission patterns}\\
The SASS count-rate as well as the source position may be uncertain for
sources which show complex emission patterns. 177 sources were flagged in the
visual process. The flag serves as a warning that the SASS count-rate and
position may be uncertain. A broad band image is available for inspection.

{\tt (v) Sources missed by the detection algorithm}\\
In the visual inspection process sources were found, which were missed by the
standard analysis software system. The number of such sources included into
the RBSC is 49. Their main parameters, count-rate, exposure time, and position
were determined in an interactive process using EXSAS tools.

\subsection{Statistical properties}

\subsubsection{Sky- and count-rate distributions}

\begin{figure}
  \centerline{\psfig{file=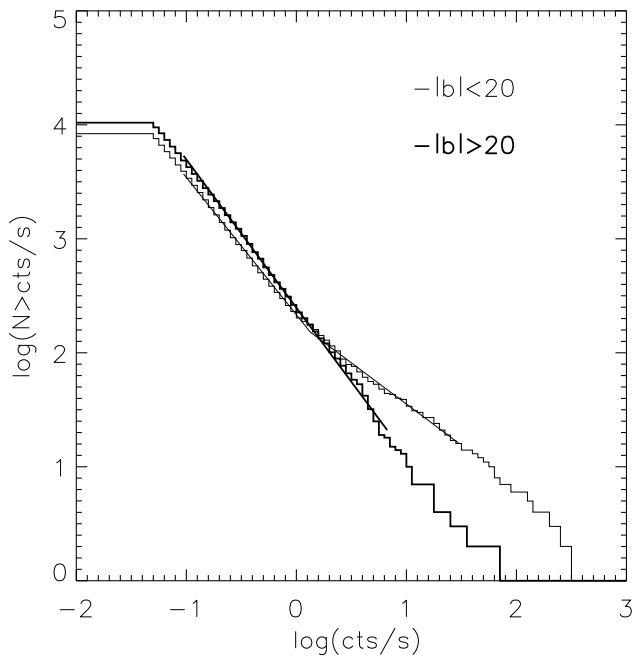,width=8.7cm,clip=}}
  \caption[Nichts]{\label{hist}
Histogram of the cumulative number density (N$>$cts/s) of RBSC sources in the
galactic plane ($\rm |b| < 20 \degr$) and outside the galactic plane ($\rm |b|
> 20 \degr$), plotted against broad band count-rate.
}
\end{figure}

\begin{figure*}
\psfig{figure=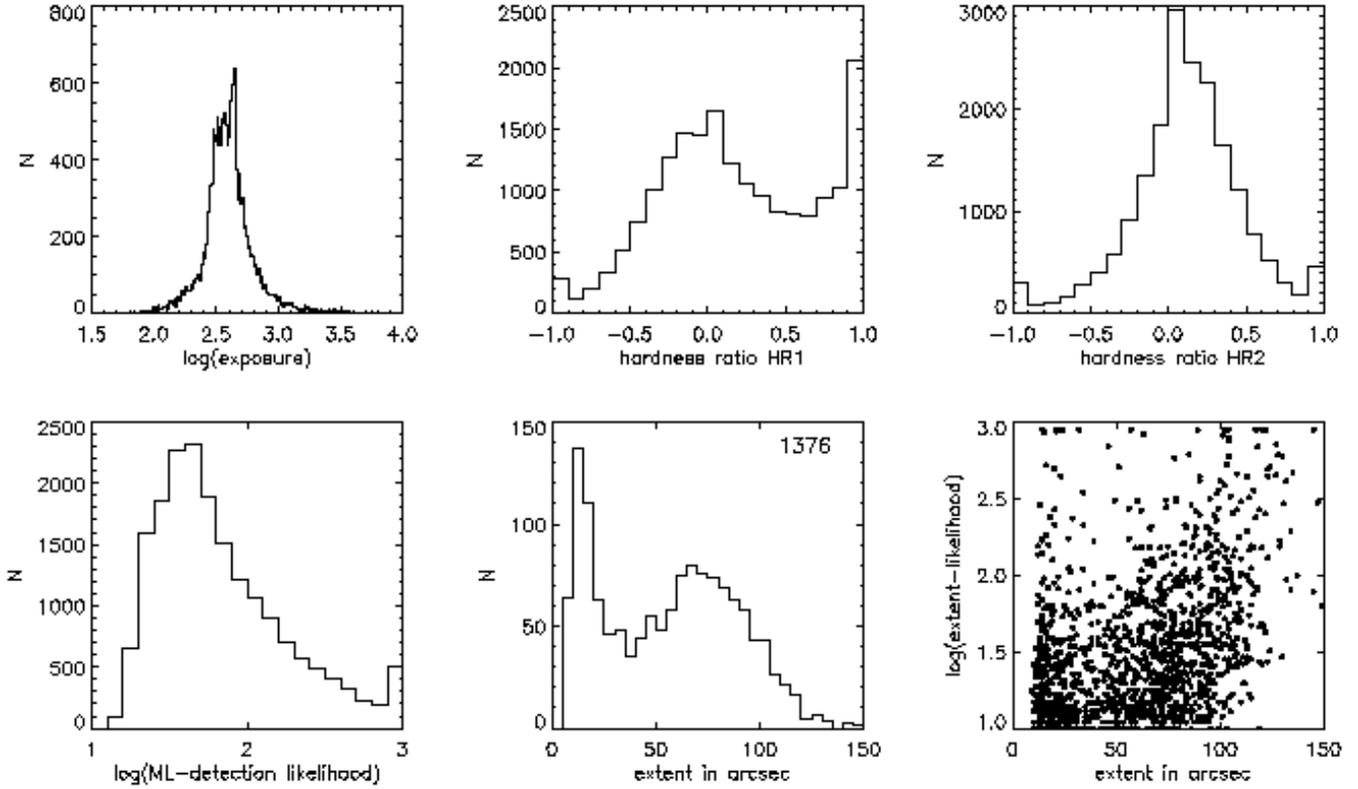,width=18.0truecm,angle=90,clip=}
\caption{
Six histograms summarizing some results of the analysis of the 18,811 RBSC
sources. The upper panel shows the distributions of exposure time,  hardness
ratios HR1 and HR2. The lower panel displays the distributions of the source
detection likelihood and the  source extent, and a plot of the extent
likelihood against extent. The last two plots are shown for sources which have
an extent likelihood $\rm \ge$ 10.
}
\end{figure*}

In Fig. 5 we present the sky distribution of all RBSC sources in galactic
coordinates. The size of the symbols scales with the logarithm of the
count-rate and the colours represent 5 intervals of the hardness ratio HR1.
The distribution of RBSC sources shows the clustering of the bright ($\rm >
1.3\ counts\ s^{-1}$) hard X-ray sources in the galactic plane, well known
from the UHURU and HEAO-1 sky surveys. At fainter count-rates ($\rm \le 1.3 \
counts\ s^{-1}$) the source distribution is more uniform. This is illustrated
in Fig. 6, where we compare the cumulative number count distributions for
sources in the galactic plane ($\rm |b| < 20 \degr$) and outside the galactic
plane ($\rm |b| \ge 20 \degr$). A linear fit to the distribution of all
sources outside the galactic plane (thick line in Fig. 6) results in a slope
of --1.30 $\pm$ 0.03. The faint line in Fig. 6 represents the count-rate
distribution for the galactic plane population of RBSC sources. The histogram
shows a break at a count-rate of about 1.3 $\rm  counts\ s^{-1}$. A linear fit
to the distribution above this break-point gives a slope of --1.20 $\pm$ 0.04;
the slope below the break-point is --0.72 $\pm$ 0.06. The flattening of the
log N--log (count-rate) of the galactic plane distribution at the bright end
is due to the disk population of bright X-ray sources (see Fig. 5).

The effect of interstellar photoelectrical absorption is demonstrated in Fig.
5 by the fact that sources near the galactic plane exhibit higher values of
the hardness ratio HR1 (blue symbols) than sources outside the galactic plane.
Although both galactic and extragalactic objects show a large spread in their
spectral energy distribution in the ROSAT band (e.g. when simple power-law
models were fit to the ROSAT spectra of broad and narrow line Seyfert 1
galaxies, the photon indices ranged between about 2 and 5), the dominant
effect here is probably the larger amount of absorption within the galactic
plane.

\subsubsection{Exposure time,  hardness ratios, detection likelihood
and source extent  distributions}

In Fig. 7 we present the distributions of some source parameters of the RBSC.
For most of the sources (see the upper left diagram) the exposure time is of
the order of a few hundred seconds. The HR1 and HR2 distributions of the RBSC
sources binned in intervals of 0.1 are shown in the upper middle and right
diagrams, respectively. This gives a more detailed representation of the HR1
distribution compared to Fig. 5, where only 5 bins were used. The distribution
of the source detection likelihood is shown in the lower left diagram. The
lower value of 15 is set by the source selection criterion for RBSC sources.
The last two diagrams show the distribution of the source extent and the
extent-likelihood as a function of extent. A more detailed discussion of the
RBSC source parameter distribution for different object classes can be found
in Sect.~4.

\subsubsection{Positional accuracy}

\begin{figure}
  \centerline{\psfig{file=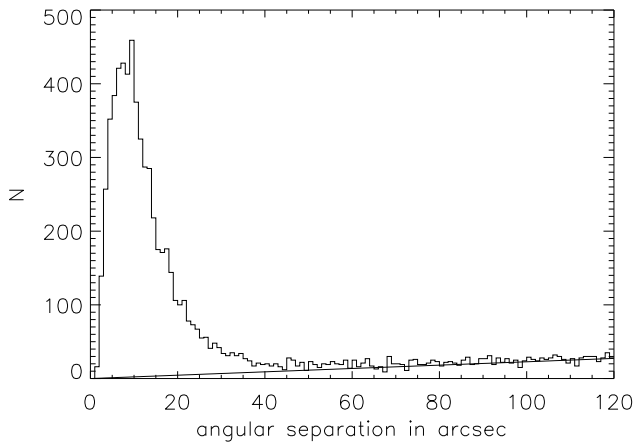,width=9.2cm,clip=}}
  \caption[Nichts]{
The plot shows the distribution of the angular separation of RBSC positions
from positions of the TYCHO catalogue for bright stars with a positional error
of less than 1 arcsec. 68\% (90\%)  of the ROSAT sources are found within 13
arcsec (25 arcsec) of the optical position. The straight line indicates the
chance coincidences, demonstrating that for distances $\geq$ 40 arcsec no
reliable ID candidates can be found.
}
\end{figure}

The RBSC sources were correlated with the TYCHO catalogue (H{\o}g et al. 1998)
to assess the positional accuracy. As TYCHO contains only stars, this
correlation gives the positional accuracy of point-like sources. Figure 8
shows the result from the correlation of the RBSC for a search radius up to
120 arcsec with the TYCHO catalogue entries. The comparison shows that 68\%
(90\%) of the RBSC sources are found within 13 arcsec (25 arcsec) of the
optical position.

\subsubsection{Temporal variability of RBSC sources}

We have investigated the temporal variability of RBSC sources on time scales
of months to years by comparing the ROSAT All-Sky Survey observations with 
public pointed ROSAT PSPC observations (see also Voges \& Boller 1998). The
comparison was done  with a search radius of 60 arcsec around the RBSC source
position. The resulting number of RBSC sources which have counterparts in
ROSAT public pointed observations is  2,611. Figure 9 shows the ROSAT survey
count-rate versus the pointing count-rate for these sources. Sources showing a
factor of variability above 5 (109), were visually inspected, similar to the
screening process performed for the RBSC sources described in Section 3.2, to
ensure the reliability of the source existence and of the source count-rate.
20 RBSC sources exhibit a factor of  variability between 10 and 100. 5 RBSC
sources show a factor of variability above 100. There is an excess of sources
with factors of variability above 10, which are brighter during the RASS
observations. This is most probably due to the fact, that during the RASS
observations the count-rate threshold for the source detection is on average
higher with respect to ROSAT pointed observations. As a result, the RBSC
sources shown in Fig. 9 are biased towards sources with extreme variability. A
comprehensive variability analysis of ROSAT sources will be presented
elsewhere.

\begin{figure}
  \centerline{\psfig{file=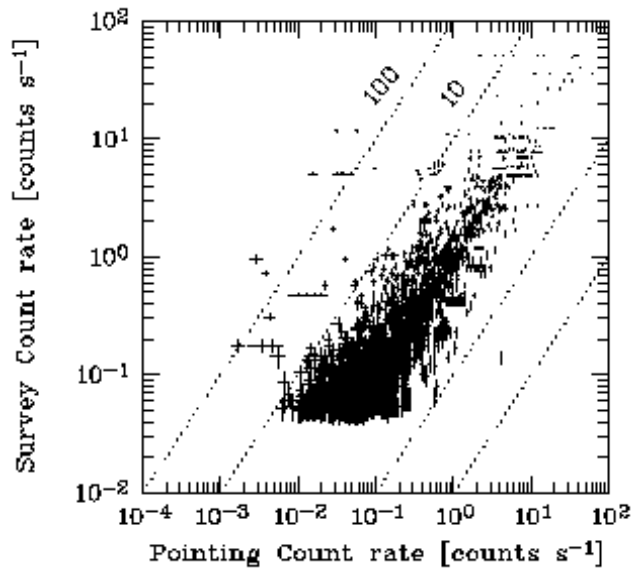,width=8.5cm,clip=}}
  \caption[Nichts]{
Count-rate of RBSC sources versus count-rate of corresponding 
pointed PSPC observations (0.1--2.4 keV).
Dashed lines are for fixed factors of variability. 
}
\end{figure}

\subsubsection{Flux determination and log N$-$log S distributions}

In order to facilitate the use of the catalogue for statistical studies it may
be useful to quote not only count-rates but photon fluxes. To convert
count-rates into fluxes in the 0.1--2.4 keV energy range we have used two
different models: Model 1 assumes a power law $\rm E^{-\Gamma + 1}\ dE$ and
may be useful for AGN and clusters of galaxies. We use a fixed photon index of
$\rm \Gamma = 2.3$, which is the typical value derived from ROSAT observation
of extragalactic objects (see Hasinger et al. 1991, Walter \& Fink 1993), and
an absorbing column density fixed at the galactic value along the line of
sight (Stark et al. 1992). These fluxes, corrected for galactic absorption,
are called flux1. Model 2 is based on an empirical conversion between
count-rates and fluxes following Schmitt et al. (1995), originally developed
to obtain flux values for stars: $$\rm flux2 = (5.3 \cdot HR1 + 8.31) \cdot
10^{-12} \cdot counts\ s^{-1} [erg\ cm^{-2}\ s^{-1}]$$ Both flux values are
listed in the electronic form of our catalogue.

For a statistical study we use both fluxes for three types of objects, stars
from the TYCHO catalogue, clusters of galaxies (ACO) from the compilation of
Abell et al. (1989)  and AGN listed in the Veron catalogue. We have further
subdivided the samples in two categories; A: point sources having an
extent-likelihood value of zero; B: sources with an extent-likelihood $\rm > $
0.

\begin{figure}
\psfig{figure=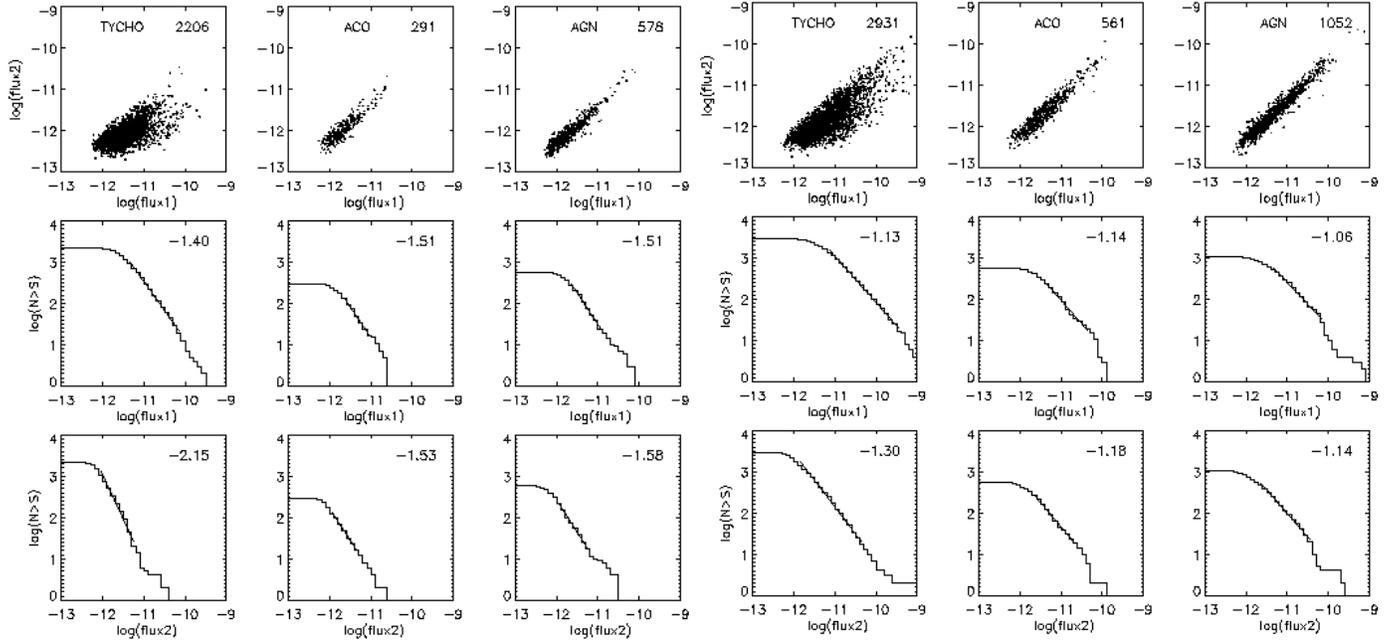,width=9.1truecm,angle=0,clip=}
  \caption[Nichts]{
Nine histograms summarizing the comparison of two different methods for the
determination of fluxes and logN$-$logS distributions for 
point sources (extent-likelihood = 0); first column: TYCHO-stars, second
column: clusters of galaxies, third column: active galactic nuclei.
The upper panel shows the correlation of flux2 with flux1,
the middle panel shows the logN -- log(flux1) distribution, and the lower 
panel shows the logN -- log(flux2) distribution.
The number of sources and fitted slope values are inserted.
}
\end{figure} 

\begin{figure}
\psfig{figure=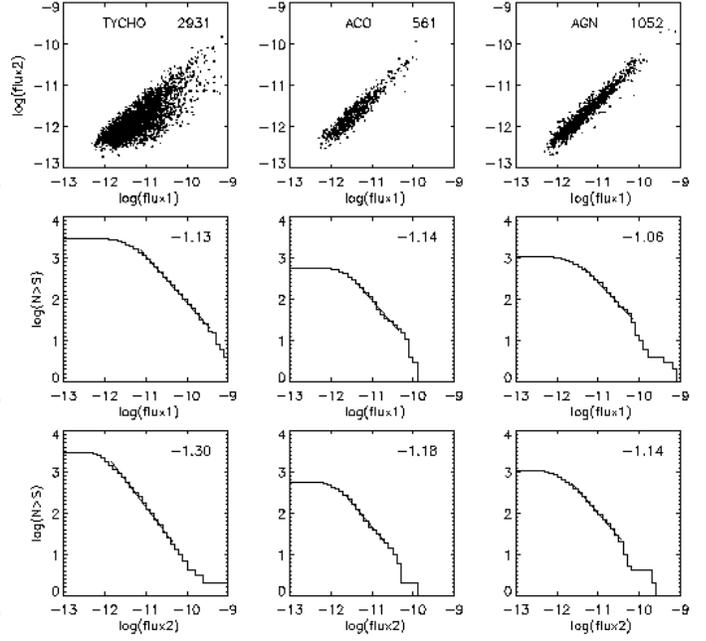,width=9.1truecm,angle=0,clip=}
  \caption[Nichts]{
Nine histograms summarizing the comparison of two different methods for the
determination of fluxes and logN$-$logS distributions for 
point-like and extended sources (extent-likelihood $\rm > $ 0); 
first column: TYCHO-stars, second
column: clusters of galaxies, third column: active galactic nuclei.
The upper panel shows the correlation of flux2 with flux1,
the middle panel shows the logN -- log(flux1) distribution, and the lower 
panel shows  the logN -- log(flux2) distribution.
The number of sources and fitted slope values are inserted.
}

\end{figure}

In Fig. 10 we compare the two different flux determinations for the three
object classes mentioned above which fall into category A. For TYCHO stars we
obtain the largest dispersion between the two methods (left plot in the upper
panel). The differences in fluxes are most likely due to the correction for
the galactic absorption. The first method tends to overestimate the flux, as
the galactic hydrogen column density is taken along the full line of sight.
The true value is lower than the value taken in the computation of flux1. For
stars the flux determination with method 2 is more reliable than method 1. A
much smaller dispersion is found in comparing flux 1 and flux 2 for ACO and
AGN. This is probably explained by the fact that both source populations are
detected at high galactic latitudes, where the amount of galactic absorption
is considerably smaller than in the galactic plane.

In Fig. 11 we compare the two different flux methods for the three object
classes which fall into category B. Again, for TYCHO stars we obtain the
largest dispersion between the two methods (left plot in the upper panel). The
differences in fluxes are due most likely to the correction for the galactic
absorption. As in Fig. 10, a smaller dispersion is found for ACO and AGN
objects.

The log N$-$log S distributions for the three object classes of category A are
shown in the middle panel (using flux1 values) and in the lower panel (using
flux2 values) of Fig. 10. We want to stress that we did not apply any
corrections concerning the varying sensitivity of the survey to the log
N$-$log S distributions. To determine the slope a linear fit was made to the
distribution, starting at the turn-over point of the distribution and ending
where the cumulative number count dropped below 15. For ACO and AGN we derive
slopes of the log N$-$log S distribution close to --1.5, which is the expected
slope for an Euclidean source distribution, for both flux determination
methods. For stars the  log N$-$log S distribution is unreliable for flux1 as
discussed before. The flux2 distribution for stars (lower panel) is very steep
for category A. We have no good explanation for this steep slope. This
requires further investigation.

The log N$-$log S distribution for the three object classes of category B are
shown in the middle (using flux1 values) and lower panels (using flux2 values)
of Fig. 11. The log N$-$log S distributions and their slopes for ACO and AGN
objects are comparable with each other for both object types and both flux
determination methods.

However, there is a general difference in the slopes for category A and
category B sources. For example, the slope for ACO is --1.51 in category A
(extl = 0) and --1.14 in category B (extl $\rm > $ 0). There are various
explanations possible for the flatter logN$-$log S distribution, such as: a)
the count-rate determination for extended sources is underestimated by the
SASS; this effect is largest for the fainter X-ray sources; b) the source
detection probability for faint extended sources decreases more rapidly than
for point sources; c) the probability to assign an extent likelihood value to
a detected source decreases with decreasing source count-rate. This effect is
stronger for faint extended sources as compared to point-like sources.

\begin{figure}
  \centerline{\psfig{file=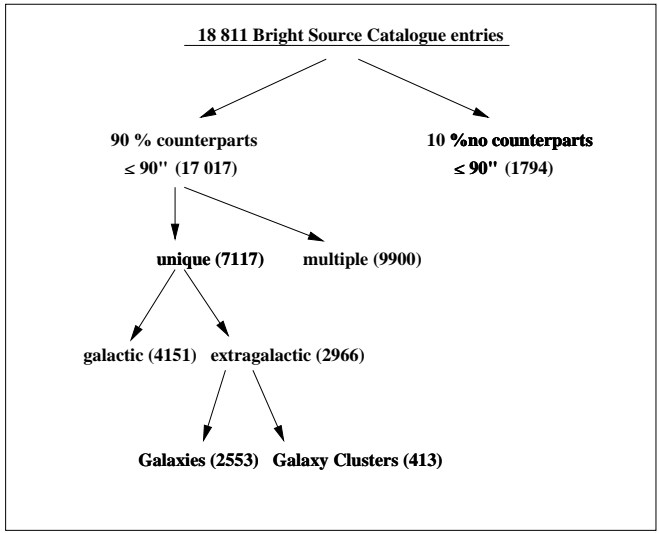,width=8.7cm,angle=-90,clip=}}
  \caption[Nichts]{\label{flux}
Diagram showing the distribution and selection of counterparts detected 
by cross-correlations with catalogues in different wavelength domains.
}
\end{figure}

Cases a,b and c are well-known software deficiencies in SASS. Various methods
have been developed by B\"ohringer et al. (1998) (growth curve method),
DeGrandi et al. (1997) (steepness ratio technique), Ebeling et al. (1993)
(Voronoi, tesselation and percolation method), 
and Wiedenmann et al. (1997) (Scaling index method), 
to attack this problem, in particular for the study of clusters of galaxies.

\section{Correlation with existing data bases}

We have performed a cross-correlation of the RBSC with various catalogues.
These catalogues include public data bases like the NED or SIMBAD and the
latest versions of published catalogues or catalogues in print, as well as
lists which were made available to us by private communications from the
following alphabetically listed authors: Alcala et al. (1995, 1996, 1998)
(T-Tauri stars), Appenzeller et al. (1998) (optical identifications of
northern RASS sources), Bade et al. (1998) (The Hamburg/RASS Catalogue of
optical identifications), Bergh\"ofer et al. (1996) (OB-stars), Beuermann et
al. (1999) (identification of soft high galactic latitude RASS X-ray sources),
Boller et al. (1998) (IRAS galaxies), Buckley et al. (1995), Burwitz et al.
(1996a, 1996b) (individual stellar and cataclysmic variables candidates),
Covino et al. (1997), Danner (1998) (stellar candidates in star forming
regions), Fleming (1998), Fleming et al. (1995, 1996), (EINSTEIN extended
medium sensitivity survey as well as white dwarf and M dwarf detections),
Gregory \& Condon (1991), Gregory et al. (1996) (radio sources), Haberl et al.
(1994), Haberl \& Motch (1995) (intermediate polars), Hakala et al. (1993)
(cataclysmic variable sources), Harris et al. (1994) (EINSTEIN catalogue of
IPC sources), Schwope et al. (1999) (identifications
of RBSC sources), Hoffleit \& Warren (1991) (WFC Bright source catalogue),
H\"unsch et al. (1998a, 1998b, 1999) (bright and nearby giant, subgiant and
main-sequence stars), Kock (1998), Kock et al. (1996), Krautter et al. (1997),
Kunkel (1997) (T-Tauri stars), Laurent-Muehleisen et al.(1998) (spectroscopic
identification of AGN in the RASS-Green Bank catalogue), Law-Green et al.
(1995), Magazzu et al. (1997) (T-Tauri stars), Mason et al. (1992) (white
dwarfs), Metanomski et al. (1998) (photometry of F, G and K stars in the
RASS), Motch et al. (1996, 1997a, 1997b, 1998) (Galactic plane survey), Nass
et al. (1996) (Hamburg catalogue of bright source identifications),
Neuh\"auser et al. (1995, 1997) (T-Tauri stars), Perlman et al. (1996) (BL
Lacertae objects), Romer et al. (1994) (clusters of galaxies), Schmitt (1997)
(A, F and G stars in the solar vicinity with distances less than 13 pc),
Schmitt et al. (1995) (K and M stars in the solar vicinity with distances
less than 7 pc), Staubert et al. (1994) (cataclysmic variable sources), Stern
et al. (1995) (RASS identifications in the Hyades cluster), Thomas et al.
(1996, 1998) (optical identification of RASS sources), Wagner et al. in 
preparation
(BL Lacertae objects), Wei et al. in preparation (quasars), 
Wichmann et al. (1996,
1997) (T-Tauri stars), Xie et al. (1997) (active galactic nuclei). These
catalogues result in significant measure from follow-up optical observations
of ROSAT sources. In these many campaigns extensive use was made of the COSMOS
(MacGillivray and Stobie (1985); Yentis et al. (1992)) and APM (McMahon \&
Irwin (1992)) digitised sky surveys.

\begin{table*}
\begin{center}
\myfontsmall{
\begin{tabular}{lrrrrrrrrrrl}
 \hline    
CATALOGUE      &  Entries & $<$300$\arcsec$ & $<$300$\arcsec$ & search  & bkg. & rad. & rad. & rad. & all& near. & References\\
               &          &  all            &near.            & rad.(sr)& cont.& 68\% & 90\% & 95.5\% &$\le$sr & $\le$sr  \\
\hline
SIMBAD         &    956370&     26931 &11687 & 82 & 9.09 & 18& 47& 65&12637& 8221& CDS, France (1998)\\
NED            & $\sim$ 800000 &38675 & 9103 &108 &17.31 & 20& 60& 86&15873& 6826& Helou et al. (1995)\\
TYCHO          &   1058332&     16737 &10274 & 42 & 3.63 & 12& 21& 28& 6156& 5465& H{\o}g et al. (1998) \\
FIRST          &    437429&      7678 & 2516 & 58 &15.74 & 16& 42& 51& 1544&  978& White et al. (1997)\\ 
NVSS           &   1814748&     23991 &11544 & 44 &12.83 & 18& 33& 39& 3594& 3437& Condon et al. (1998)\\
VERON          &     15106&      1936 & 1826 & 76 & 0.77 & 11& 20& 27& 1718& 1701& Veron-Cetty \& Veron (1998) \\
RITTER         &       414&       155 &  155 & 46 & 2.52 &  8& 17& 26&  149&  149& Ritter \& Kolb (1998) \\
PRIVATE        &     13483&     10790 & 7744 &102 & 0.17 &  9& 17& 24&10676& 7699& see Sect. 4\\
ROSATP3        &     82325&     11106 & 3631 & 86 & 5.89 & 23& 48& 65& 6677& 3300& Voges et al. (1996b) (*)\\
ROSAT-WGA      &     68907&      9451 & 3057 &120 & 7.51 & 28& 58& 80& 6624& 2908& White et al. (1994) (*)\\
ROSHRI         &      7622&      3801 &  869 &112 & 4.83 & 14& 33& 52& 2682&  844& ROSAT RRA consortium (1999) (*) \\
ROSAT-WFC      &       688&       605 &  436 & 92 & 1.23 & 24& 47& 60&  593&  429& Mason et al.(1995), Pye et al. (1995) \\  
EUVE           &       549&       322 &  319 & 98 & 4.01 & 49& 68& 80&  308&  307& Bowyer et al. (1996) \\
IRAS-FSC       &    173044&      4049 & 3711 & 62 & 3.70 & 21& 37& 48& 2205& 2200& Moshir et al. (1989)\\
IRAS-PSC       &    245889&      3923 & 3300 & 58 & 5.86 & 19& 38& 50& 1566& 1562& IRAS Cats. \& Atlases Suppl. (1988)\\
\hline 
\hline
CATALOGUE     &  Entries &    $<$60$\arcsec$ & $<$60$\arcsec$ & search  & bkg.& rad. & rad. & rad. & all& near.& References\\ 
              &          &      all         & near.           & rad.(sr)&cont.& 68\% & 90\% & 95.5\%& $\le$sr &$\le$sr\\
\hline
HST-GSC        &  25239690&     26216 & 12352 & 24 & 13.92 & 11&  18& 22&15824& 9759& Lasker et al. (1990) \\ 
             \hline
\end{tabular}}
\end{center}
      \caption{List of catalogues used in the correlation with the RBSC.
The first column gives the name of the catalogue and the second gives the
number of entries. The third and fourth columns show the number of $all$
detected counterparts and the total number of the nearest detected
counterparts, both found within a search radius of 300 arcsec ( 90 arcsec for
HST-GSC). The following two columns denote the search radii (sr, in arcsec)
used for the identifications, as well as the respective background
contamination (given in per cent) which may lead to spurious detections. The
next 3 columns give the search radii for detection at confidence levels of 68,
90 and 95.5 percent. Columns 10 and 11  present the number of $all$ detected
counterparts and the nearest neighbours within the search radius (sr). The
last column gives the references to the appropriate catalogues. NED is the
NASA/IPAC Extragalactic Database, NVSS stands for the NRAO VLA Sky Survey.
Private contributions are summarized with PRIVATE and are described in Sect. 4
and in the references. ROSATP3 denotes the ROSAT PSPC-Pointing catalogue by
Voges et al.; ROSAT-WGA is the ROSAT PSPC-Pointing catalogue by White, Giommi
\& Angelini; ROSHRI denotes the ROSAT Result Archive (RRA) catalogue of
HRI-Pointings; ROSAT-WFC summarizes the ROSAT Wide Field Camera all-sky survey
and the respective optical identifications; EUVE is the second Extreme
Ultra-Violet Explorer Catalogue. The IRAS Faint source catalogue and the IRAS
Point source catalogue are abbreviated with IRAS-FSC and IRAS-PSC,
respectively. The Hubble Guide Star Catalogue is quoted as HST-GSC. ((*)
available at http://wave.xray.mpe.mpg.de/rosat/catalogues/)
}
\end{table*}

Table 3 is a compilation of all catalogues used. In the identification process
discussed in this section we use a search radius of 90 arcsec around the RBSC
position. For 90\%  of all RBSC sources  at least one counterpart has been
found in either one of the catalogues (not taking into account ROSATP3,
ROSAT-WGA, and ROSHRI pointing catalogues), whereas 10\% have no catalogue
entries. We have divided the RBSC sources with catalogue counterparts into two
subclasses, (i) unique entries and (ii) multiple entries (see Fig.12). Unique
entries refer to RBSC sources which have only one catalogue entry in the
various catalogues, or which have a unique identification in the private
catalogues. The number of RBSC sources which have unique entries is 7,117.
Only these sources are used in the following to derive observational
properties, like the ratio of the optical- to X-ray flux. For 9,900 RBSC
sources more than one counterpart is detected within the 90 arcsec search
radius. RBSC sources with multiple entries are not included in the discussion
below. From the 7,117 unique entries 42\% are classified as extragalactic
objects and 58\% are of galactic origin. The extragalactic objects are
subdivided into galaxies, including active and non-active galaxies, and
clusters of galaxies. 2,553 RBSC sources are identified as galaxies and 413 as
galaxy clusters.

The ratio of the X-ray-to-optical flux is a useful discriminant of the nature
of the X-ray source, in particular in discriminating between stars and
extragalactic objects. Figure 13 shows a correlation of the flux ratio with
the optical magnitude for a sample of RBSC sources already identified as
stars, AGN, or clusters of galaxies. The gap between stars and AGN/ACO is
partly an artifact as star catalogues become incomplete at $ m_v > 12$.

\begin{figure*}

\psfig{figure=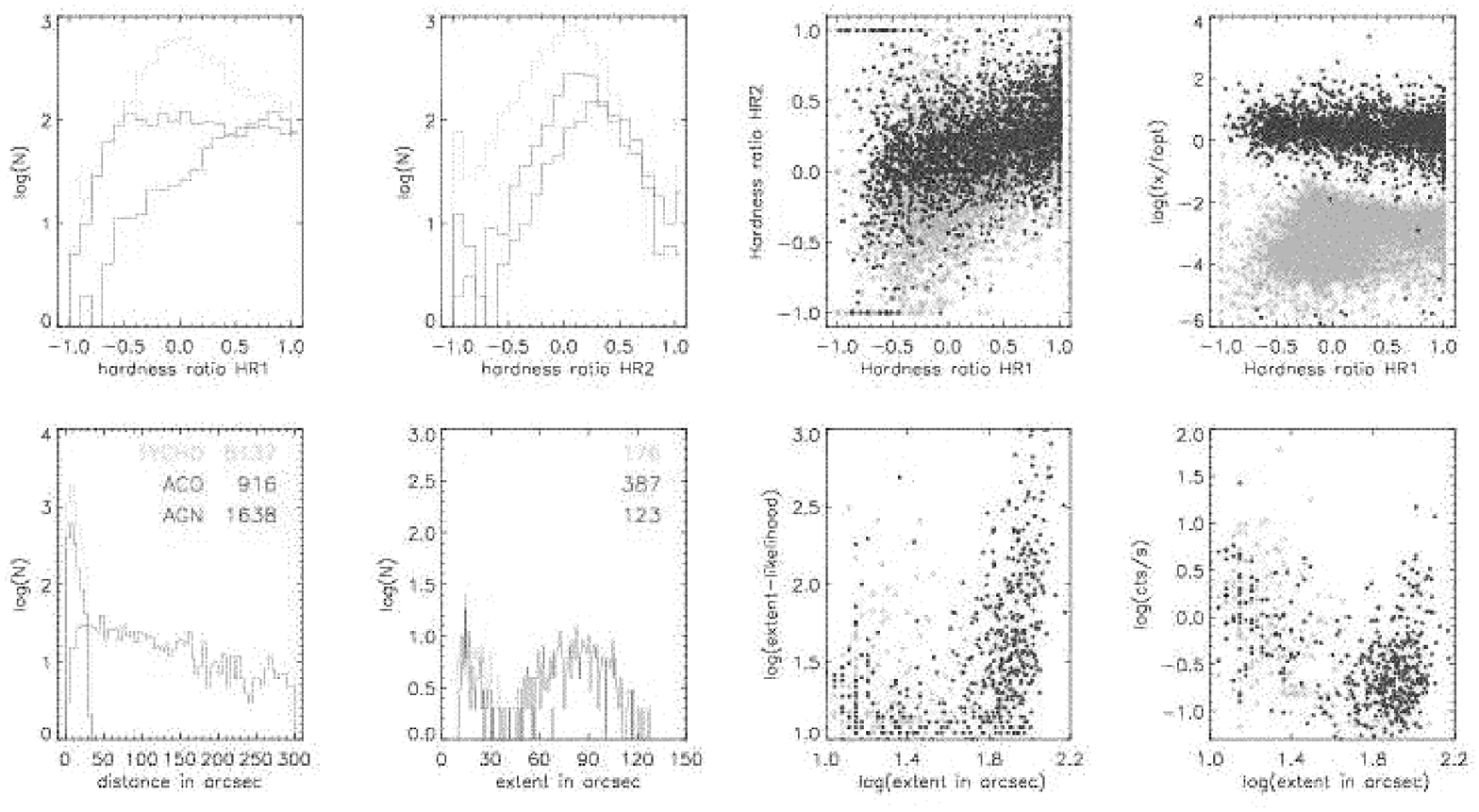,width=18.0truecm,height=8.6cm,angle=90,clip=}
\setcounter{figure}{13}
\caption{
    Eight histograms summarizing some results of the analysis of the RBSC
    sources for 5,137 TYCHO stars (in green), 916 Abell clusters of galaxies (ACO)
    (in blue), and 1,638 active galactic
    nuclei and quasistellar objects of the VERON catalogue (AGN) (in red).
    The upper panel shows the distributions of the spectral hardness
    ratios HR1 and HR2, correlations of HR2 with HR1, and $\rm f_X/f_{opt}$ with
    HR1.
    The lower panel shows the distribution of the angular separation of RBSC
    positions from the optical positions,
    source angular extent, correlations of the extent likelihood
    and the count-rate with extent.
    The last three plots are shown for sources which have an
    extent likelihood $\rm \ge $ 10.}
\vspace*{2ex}
\begin{minipage}[t]{8.7cm}
\centerline{\psfig{file=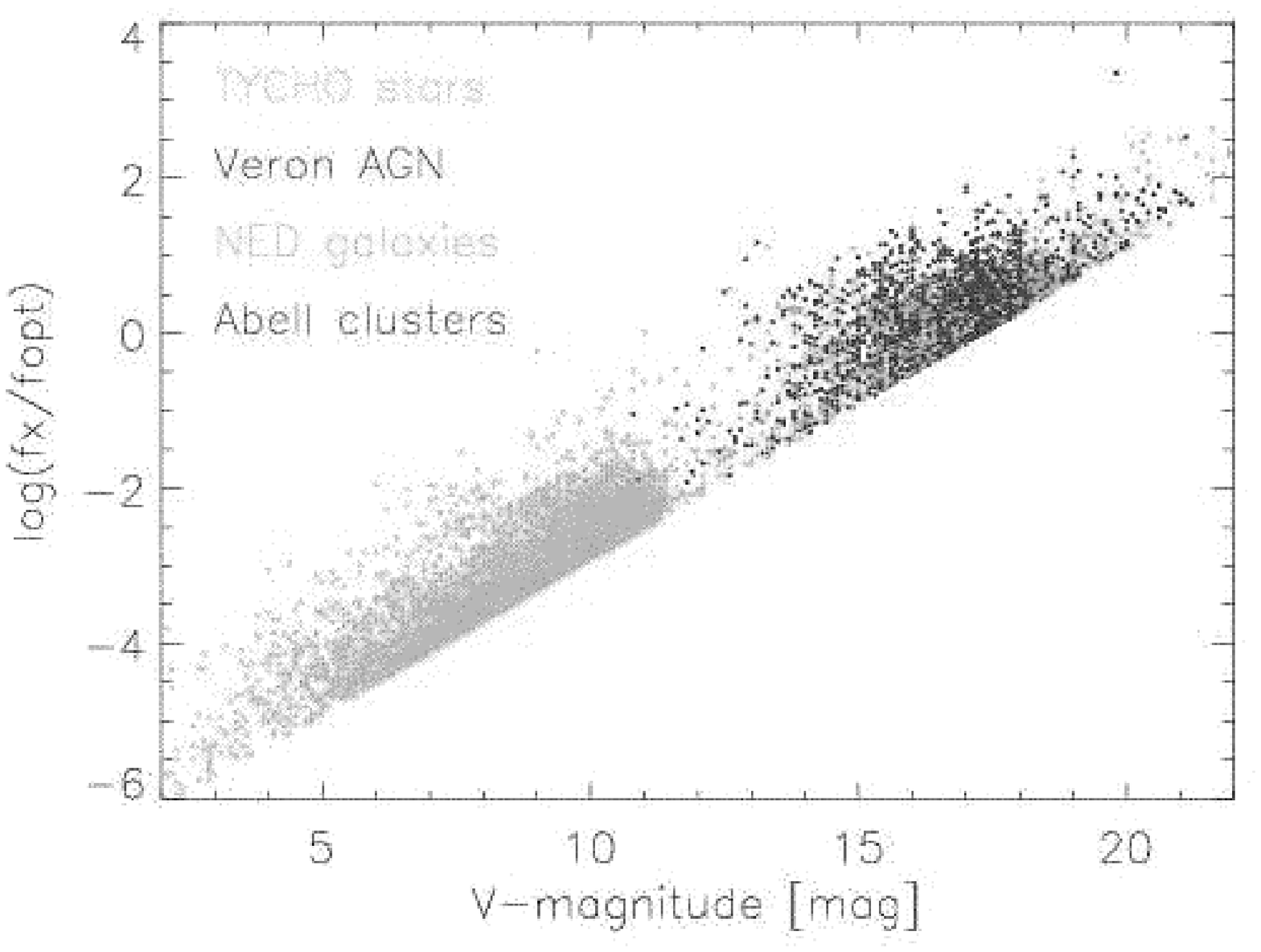,width=8.7cm,angle=90,clip=}}
\setcounter{figure}{12}
\caption[Nichts]{
    Ratio of X-ray to optical flux $\rm f_X/f_{opt}$ versus optical magnitude
    $m_{V}$ for RBSC sources with unique identification candidates.
    $\rm f_X/f_{opt}$
    was calculated from the 0.1$-$2.4 keV count-rate
    and $m_{V}$:
    $\rm log(f_X/f_{opt}) = log(PSPC\ counts/s \cdot  10^{-11}) + 0.4\ m_V + 5.37$
    (Maccacaro et al. 1988).
    Objects with unique identification are colour-coded: stars (green),
    active galactic nuclei (red), galaxies (yellow) and clusters of galaxies (blue).
    Due to the X-ray count-rate limit of 0.05 $\rm counts\ s^{-1}$ (flux limit of
    $\rm 5 \cdot  10^{-13}\ erg\ cm^{-2}\ s^{-1}$) there is a sharp lower boundary
    in the distribution.}
\end{minipage}
\hspace*{0.35cm}
\begin{minipage}[t]{8.7cm}
\centerline{\raisebox{1.5cm}{\psfig{file=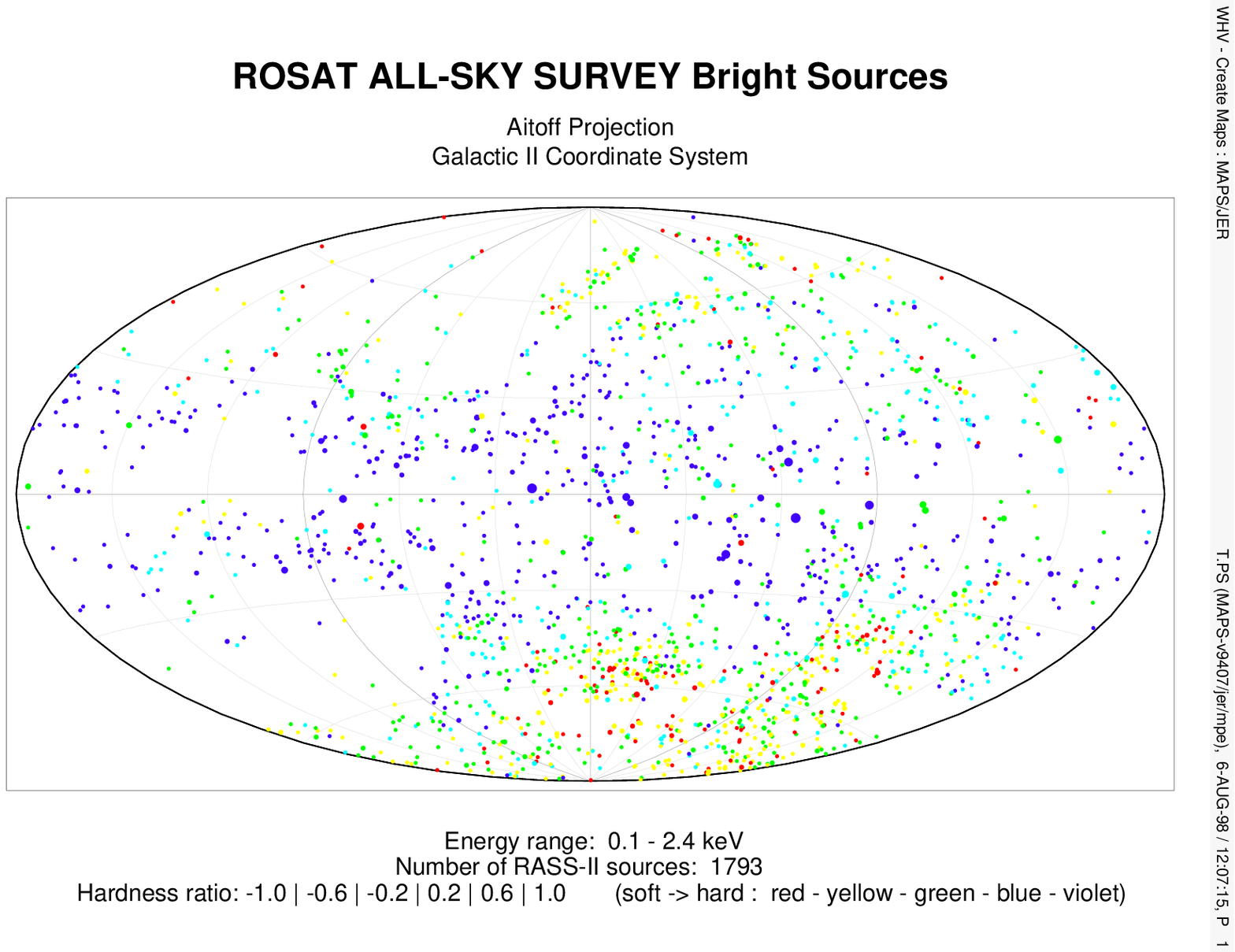,width=8.7cm,angle=-90,clip=}}}
\vspace*{1ex}
\setcounter{figure}{14}
\caption[Nichts]{
    All-sky distribution in galactic coordinates (Aitoff projection)
    of RBSC sources
    which have no identification within a search radius of 90 arcsec,
    using the catalogues of Table 3.
    The size of the symbols scales with the logarithm
    of the count-rate and the colours represent 5 intervals of the
    hardness ratio HR1:
    red ($\rm -1 \le HR1 < -0.6$);
    yellow ($\rm -0.6 \le HR1 < -0.2$);
    green ($\rm -0.2 \le HR1 < 0.2$);
    blue ($\rm 0.2 \le HR1 < 0.6$) and
    violet ($\rm 0.6 \le HR1 \le  1.0$).
    }
\end{minipage}

\end{figure*}

\begin{figure*}
\centerline{\psfig{file=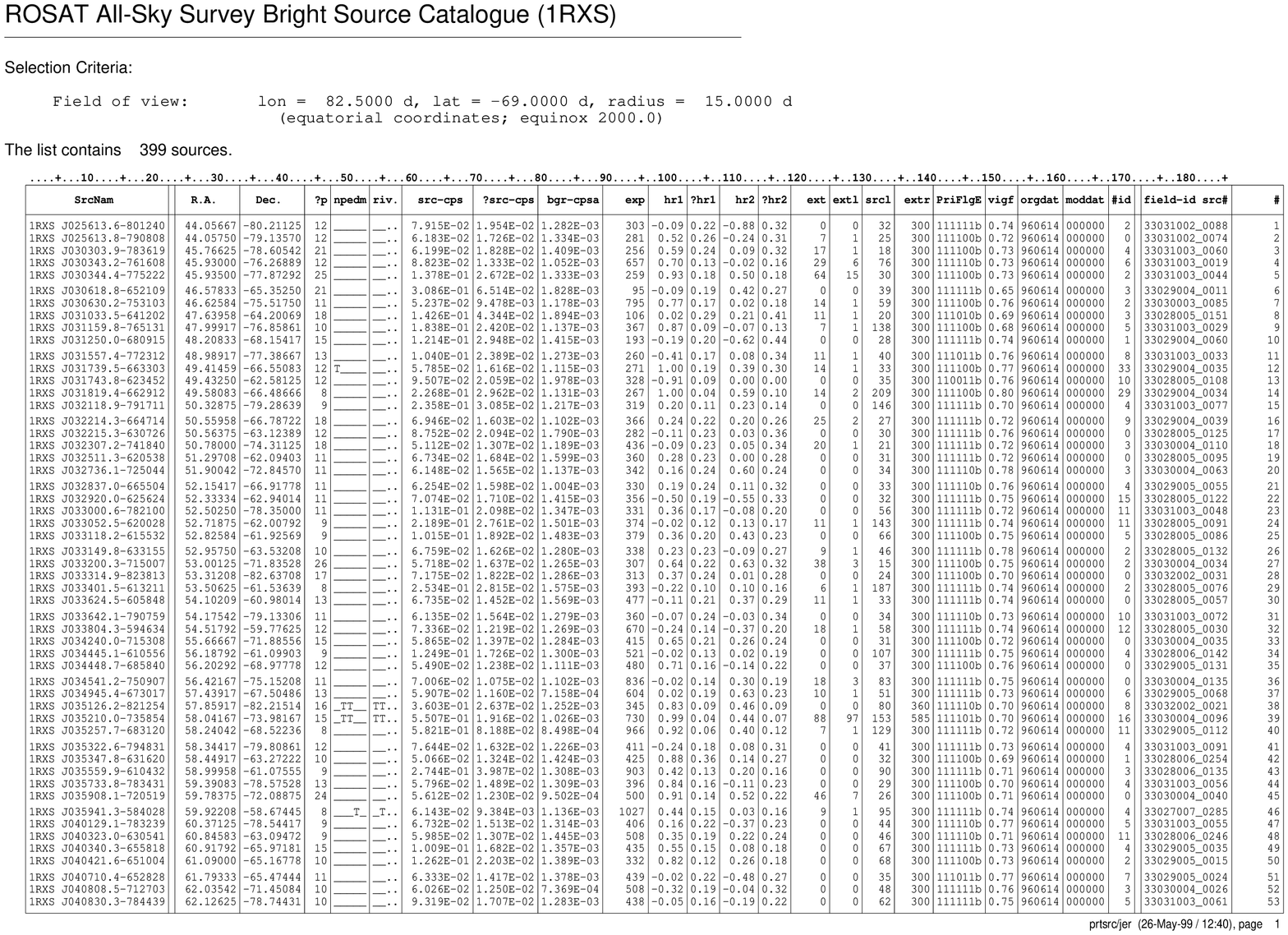,height=5.0cm,width=18.0cm,angle=-90,clip=}}
\vspace*{0.5ex}
Description of the columns:
\begin{center}{\scriptsize
\begin{tabular}{rlp{14.7cm}}
 \hline 
  \struth Col & abbreviation &  description   \\
 \hline
\strutt
  1  & SrcNam  &  the ROSAT All-Sky Survey Catalogue source name        \\
  2  & R.A.    &  Right Ascension (2000) in decimal degrees             \\
  3  & Dec.    &  Declination (2000) in decimal degrees                 \\
  4  & ?p      &  the total positional error ($1 \sigma$--radius) in arcsec,
                     including 6'' systematic error                           \\
  5  & npedm   &  screening flags ('T' for 'true', 'F' or '-'
                     for 'false', '.' for 'not used') with the following
                     denomination:
                     {\bf n} for nearby sources affecting SASS flux 
                        determination,
                     {\bf p} for possible problem with position 
                        determination,
                     {\bf e} for source extended beyond SASS 
                        extraction radius,
                     {\bf d} for complex diffuse emission pattern,
                     {\bf m} for source missed by SASS                        \\
  6  & riv.    &  additional flags which are abbreviated with:
                     {\bf r} source counts and extraction radius 
                        recalculated,
                     {\bf i} for broad band image available,
                     {\bf v} for variability flag (not yet filled);
                     {\bf .} is a dummy flag                                  \\
  7  & src-cps &  source count-rate (vignetting corrected) in the broad band in
                   counts/sec     \\
  8  & ?src-cps&  error of source count-rate in the broad band
                     in counts/sec                                            \\
  9  & bgr-cpsa&  background count-rate (vignetting corrected)
                     in counts/sec/arcmin$^2$                                 \\
 10  & exp     &  exposure time in sec                                  \\
 11  & hr1     &  hardness ratio 1 (as explained in Sect. 2.2.3.)       \\
 12  & ?hr1    &  error of hardness ratio 1                             \\
 13  & hr2     &  hardness ratio 2 (as explained in Sect. 2.2.3.)       \\
 14  & ?hr2    &  error of hardness ratio 2                             \\
 15  & ext     &  source extent in arcsec                               \\
 16  & extl    &  likelihood of source extent                            \\
 17  & srcl    &  likelihood of source detection defined as
                     srcl = --ln (1 -- P) (P = probability of source existence) \\
 18  & extr    &  extraction radius in arcsec  \\
 19  & PriFlgE &  priority flags derived from the sliding window
                     detection history using either the background map (M)
                     or the local background (B) where 0 = no detection
                     and 1 = detection with the order of flags:
                     M--broad, L--broad, M--hard, L--hard, M--soft, L--soft;
                     E indicates the PHA range with highest detection likelihood
                     (as explained in Sect. 2.2.2.)      \\
 20  & vigf    &  vignetting factor \\
 21  & orgdat  &  date (yymmdd) when the source was included \\
 22  & moddat  &  date (yymmdd) when the source properties were changed \\
 23  & \# id   &  number of possible identification objects found within
                        a search radius of 5 arcmin                    \\
 24  & field-id src\# & the identification number of the SASS field,
                   and the SASS source number                          \\
\strutb
 25  & \#      &  running source number                                \\
\hline
\end{tabular}
}\end{center}
%
%
\vspace*{-3ex}
\caption[]{
   Truncated sample output of a RBSC search inquiry
   with indicated coordinates and search radius
   (via the ROSAT source browser).
}
\end{figure*}

In Fig. 14 we compare results of the analysis of RBSC sources for Tycho stars,
clusters of galaxies and active galactic nuclei. The distribution of ACO
objects show a strong increase towards larger HR1 values, i.e. the majority of
ACO objects are much harder as compared to stars and AGN. The AGN show a flat
distribution in the range between --0.5 to +1.0. Below HR1=$-$0.5 the
distribution drops quite rapidly due to galactic absorption and due to an
intrinsic dispersion in the slopes of the X-ray continua.  The HR1
distribution for stars peaks at HR1 = 0. Most of the stars are found in the
range --0.5 to +0.5. The histograms of the hardness ratio HR2 confirm the
nature of ACO clusters as ``hard" sources. The distribution for stars and AGN
are softer and alike. In the distribution of HR2 versus HR1 ACO are dominating
the upper right region; AGN are found mostly in the central part with HR1 $>$
--0.5  and --0.5 $<$ HR2 $<$ +0.5. Stars are occupying a slightly more extended
region than AGN. The scatter plot of $\rm f_X/f_{opt}$ versus HR1 shows that
the locus of stars is well separated from AGN and ACO clusters. Stars are
primarily found in the lower portion of the plot. ACO clusters occupy only the
upper right part of the figure. The first histogram in the lower panel of Fig.
14 displays the angular separation of RBSC positions from the optical
positions. For stars and AGN most X-ray sources are found within 30 arcsec; the
distribution for clusters of galaxies is much broader as one would expect from
the X-ray source extent alone, which is shown in the next histogram. Abell et
al. (1989) determined the cluster centers visually quoting typical standard
deviations of $\pm$2'--3' for the coordinates with the positional uncertainty
depending on the compactness of the cluster. The optically determined centers
do not necessarily follow the gas distribution and thus the gravitational
potential well, which is the origin of the X-ray emission. Ebeling et al.
(1993) found that the optical position of rich clusters had a mean deviation
from the RASS position of 3' for rich clusters, and 7' for poor clusters. In
some exceptional cases angular separations of up to 15' were found. The last
two scatter plots of Fig. 14 exhibit the extent likelihood and the count-rate
versus extent. The populations of stars and AGN are well separated from those
of ACO clusters.

Figure 15 gives the celestial distribution of RBSC sources which have no
counterparts according to the catalogues listed in Table 3, when a search
radius of 90 arcsec is used. It is obvious that certain regions of the sky
have been intensively studied at other wavelengths (empty regions in Fig. 15).
In addition by comparing Fig. 15 with Fig. 5, it is evident that almost all of
the brightest RBSC sources have already been identified.

The list of possible identification candidates as generated from
cross-correlating the RBSC sources with various catalogues is intended to
assist in determining the nature of the X-ray sources. No attempt was made to
remove duplicate entries from different source catalogues and in some cases
contradictory identification candidates may be given. With the help of X-ray
parameters, such as the hardness ratios, the source extent and the $\rm
f_X/f_{opt}$ ratio, a decision can be made as to which entry is the plausible
identification candidate (also see Motch et al. 1998 and Beuermann et al.
1999).

\section{Electronic archive}

All relevant information about access to the RBSC is available at {\it
http://wave.xray.mpe.mpg.de/rosat/catalogues/ rass-bsc}. The full catalogues
(RBSC and identification catalogue) and the descriptions of the catalogue
contents can be retrieved as ASCII files via the WWW or via anonymous ftp.
The RBSC presents the X-ray data. Subsets of the catalogue sources may be
retrieved via the ROSAT Source Browser also available at the above mentioned
address. An example of a search inquiry and its output are explained in Figs.
16 -- 17.
The identification catalogue (Fig.~18 and Tab.~\ref{tdit})
consists of the results from cross-correlating the RBSC
with the catalogues listed in Tab.~3.
The main X-ray properties (such as name, position in equatorial and galactic
coordinates, and fluxes) are given, as well as a designated number which
represents the number of counterpart candidates found within a search radius
of 300 arcsec. For each of the candidates an individual record is appended,
containing the source position, angular separation from the X-ray position,
different magnitudes (or fluxes in Radio or IR wavelength bands), redshift,
spectral type and classification (if available).
The catalogues are meant to be living databases, so that whenever improved
X-ray or new ID data are available they will be included. Each data record
contains a date stamp according to year, month and day in which this record
was changed or newly included.

\begin{figure}
\centerline{\psfig{file=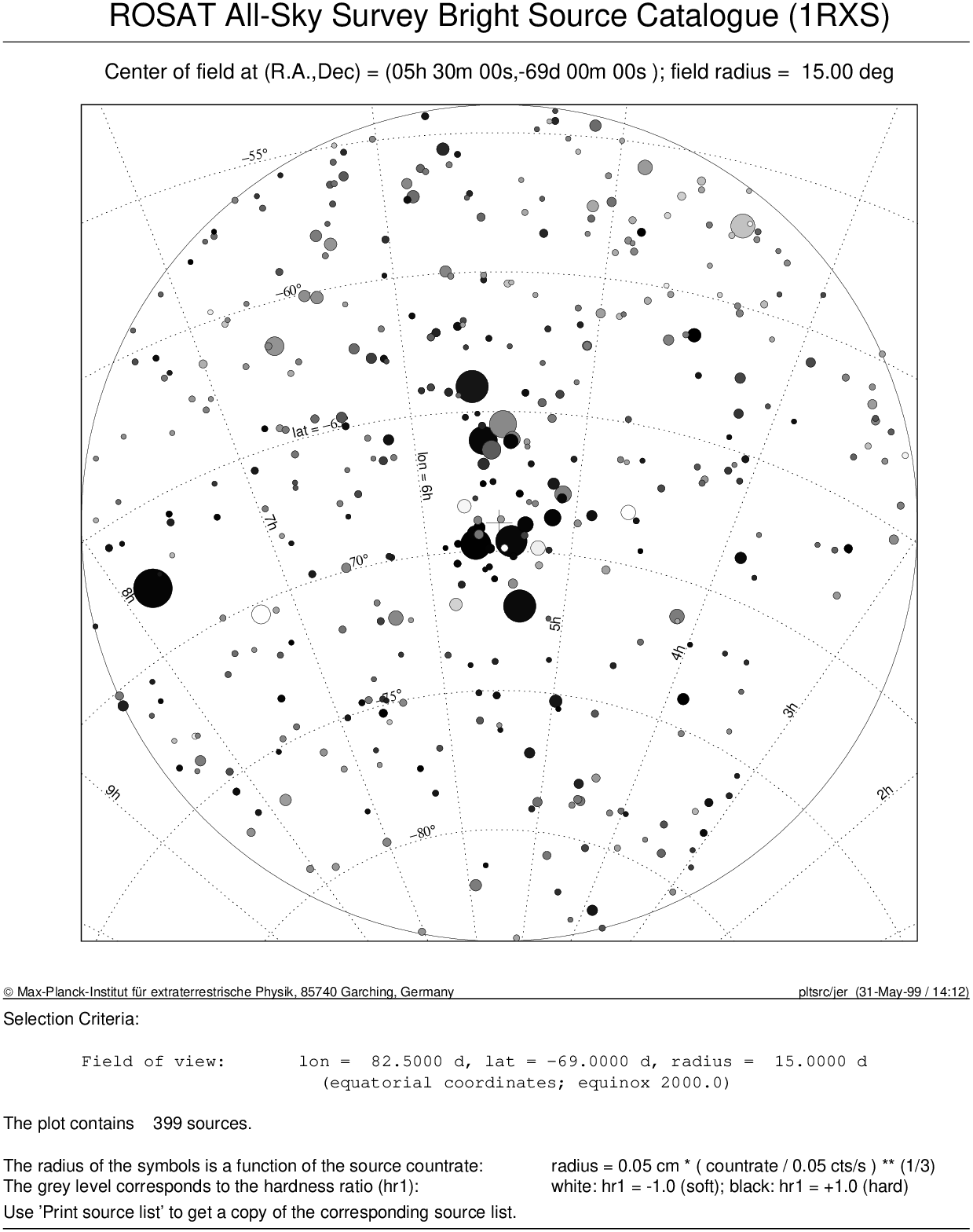,width=8.7cm,angle=0,clip=}}
\caption[]{Sample sky field plot of the search inquiry via the ROSAT source 
browser as used for Fig. 16. The size of the symbols is a function of the
count-rate and the grey level corresponds to the hardness ratio HR1.}
\end{figure}

%
\begin{acknowledgements}
We would like to thank the ROSAT team at MPE for their support and for
stimulating discussions. We thank Damir \v{S}imi\'c  for help in compiling the
identification data and Ray Cruddace for the critical reading of the
manuscript. The ROSAT Project is supported by the Bundesministerium f\"ur
Bildung und Forschung (BMBF/DLR) and the Max-Planck-Gesellschaft (MPG). This
work has made use of the SIMBAD database operated at CDS, Strasbourg, France.
In addition this research also made use of the NASA/IPAC Extragalactic
Database (NED), which is operated by the Jet Propulsion Laboratory, California
Institute of Technology, under contract with NASA, and the COSMOS digitised
optical survey of the southern sky, operated by the Royal Observatory
Edinburgh and the Naval Research Laboratory, with support from NASA. We thank
Richard MacMahon for making the APM catalogue availabe to MPE.

\end{acknowledgements}
%

\begin{figure*}
\centerline{\fbox{\psfig{file=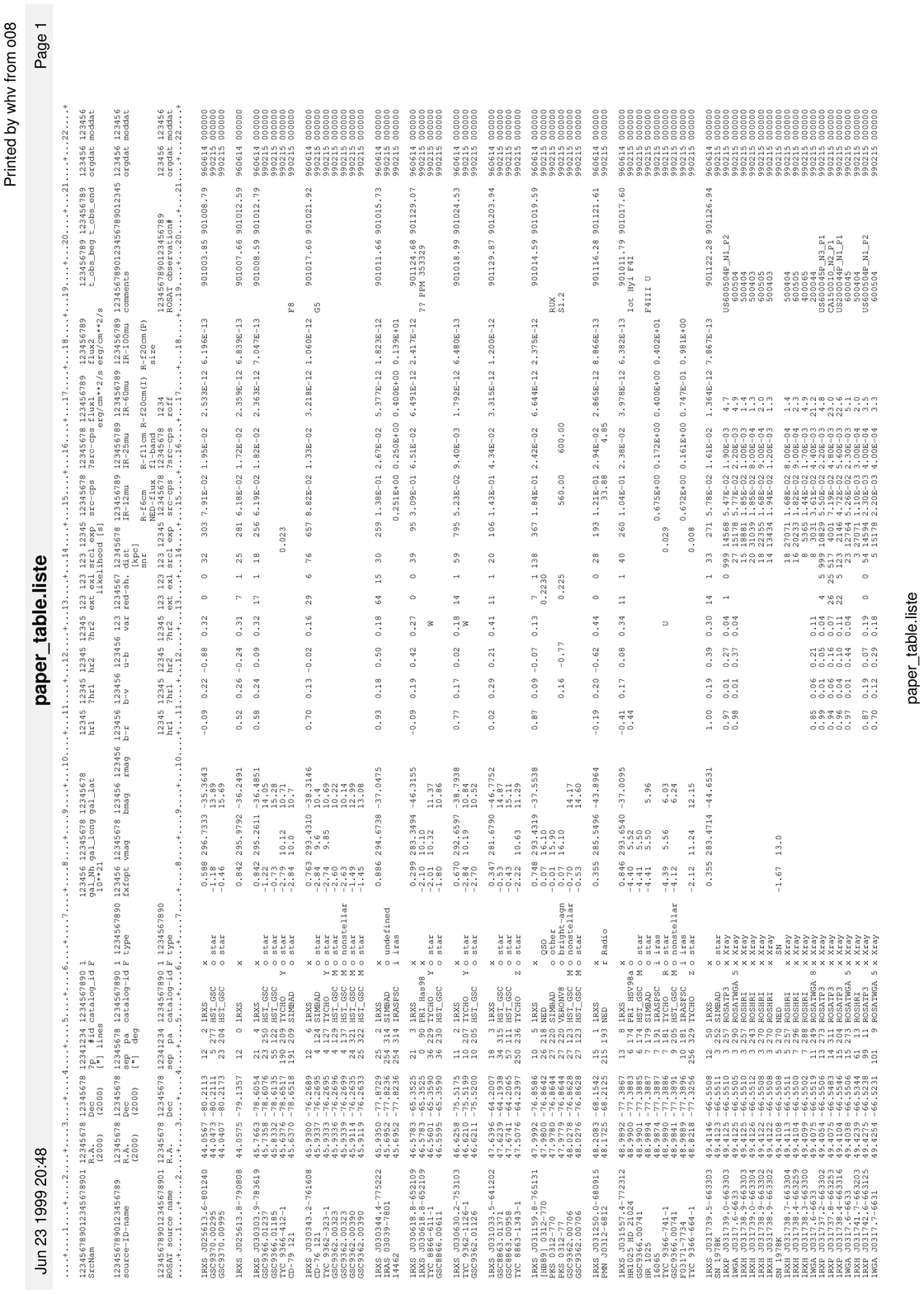,height=22.0cm,angle=0,clip=}}}
  \caption[]{
Truncated sample output of the identification table containing results from 
cross-correlating source positions of Fig. 16 with the catalogues 
listed in Tab. 3. Each block consists of one line with information
pertaining to a RBSC source followed by Nid lines containing information
from the various catalogues sorted according to increasing positional source
separation; 
see table~\ref{tdit} for a description of the columns of the 
identification table.
}
\end{figure*}

\begin{table*}

\caption{\label{tdit} Description of the columns of the identification table}

RBSC row:
\vspace*{0.5ex}
\begin{center}{\scriptsize
\begin{tabular}{rlp{14.7cm}}
 \hline 
  \struth Col & abbreviation &  description   \\
 \hline
\strutt
  1  & SrcNam     &  the ROSAT All-Sky Survey Catalogue source name     \\
  2  & R.A.       &  Right Ascension (2000) in decimal degrees          \\
  3  & Dec.       &  Declination (2000) in decimal degrees              \\
  4  & ?p         &  the total positional error ($1 \sigma$--radius) in arcsec,
                        including 6'' systematic error        \\
  5  & \# id      &  number of possible identification objects found within
                        a search radius of 5 arcmin 
                        (= number of ID-rows following)                       \\
  6  & catalog-id &  name of source catalogue (always 1RXS)                   \\
  7  & F          &  flag indicating the origin of the source position
                        (always x: X-ray)                                     \\
  8  & gal\_Nh    &  galactic HI column density in units 
                        of $\rm 10^{21} atoms / cm^2 $                  \\
  9  & gal\_long  &  galactic longitude l$^{\rm II}$ in decimal degrees     \\
 10  & gal\_lat   &  galactic latitude b$^{\rm II}$ in decimal degrees      \\
 11  & hr1        &  hardness ratio 1 (as explained in Sect. 2.2.3.)    \\
 12  & ?hr1       &  error of hardness ratio 1                          \\
 13  & hr2        &  hardness ratio 2 (as explained in Sect. 2.2.3.)    \\
 14  & ?hr2       &  error of hardness ratio 2                          \\
 15  & ext        &  source extent in arcsec                            \\
 16  & exl        &  likelihood of source extent                        \\
 17  & srcl       &  likelihood of source detection defined as          
                     srcl = --ln P (P = probability that source is a
                     spurious detection)                                \\
 18  & exp        &  exposure time in sec                               \\
 19  & src-cps    &  source count-rate (vignetting corrected) in the broad band
                        in counts/sec      \\
 20  & ?src-cps   &  error of source count-rate in the broad band
                        in counts/sec      \\
 21  & flux1      &  X-ray flux 1 in the 0.1--2.4 keV energy range
                        in $\rm erg\ cm^{-2}\ s^{-1}$
                        (as explained in Sect. 3.3.5.) \\
 22  & flux2      &  X-ray flux 2 in the 0.1--2.4 keV energy range
                        in $\rm erg\ cm^{-2}\ s^{-1}$
                        (as explained in Sect. 3.3.5.) \\
 23  & t\_obs\_beg  &  first date (yymmdd.dd) of observation
                        (.dd is the fraction of the day)     \\
 24  & t\_obs\_end  &  last date (yymmdd.dd) of observation
                        (.dd is the fraction of the day)     \\
 25  & orgdat     &  date (yymmdd) when the source was included \\
\strutb
 26  & moddat     &  date (yymmdd) when the source properties were changed \\
\hline
\end{tabular}
}\end{center}

Identification row: 
\vspace*{0.5ex}
\begin{center}{\scriptsize
\begin{tabular}{rlp{14.7cm}}
 \hline
  \struth Col & abbreviation &  description   \\
 \hline
\strutt
  1  & source-ID  &  the source name                                    \\
  2  & R.A.       &  Right Ascension (2000) in decimal degrees          \\
  3  & Dec.       &  Declination (2000) in decimal degrees              \\
  4  & sep        &  the positional separation between RBSC source
                        and ID object in arcsec                               \\
  5  & pa         &  the position angle (North to East) between RBSC source
                        and ID object in degrees                              \\
  6  & catalog-id &  name of source catalogue                                 \\
  7  & F          &  flag indicating the origin of the source position
                        (o: optical; x: X-ray; r: Radio; e: EUV; i: IR)       \\
  8  & type       &  type or class of object                                  \\
  9  & fxfopt     &  ratio of X-ray to optical flux $\rm f_X/f_{opt}$         \\
 10  & vmag       &  visual magnitude                                         \\
 11  & bmag       &  blue magnitude                                           \\
 12  & rmag       &  red magnitude                                            \\
 13  & b-r        &  (blue--red) colour                                         \\
     & hr1        &  {\bf or} hardness ratio 1 if entry is from ROSATP3 or ROSATWGA  catalogue                    \\
 14  & b-v        &  (blue--visual) colour                                      \\
     & ?hr1       &  {\bf or} error of hardness ratio 1
                        if entry is from ROSATP3 or ROSATWGA catalogue                    \\
 15  & u-b        &  (ultraviolet--blue) colour                               \\
     & hr2        &  {\bf or} hardness ratio 2
                        if entry is from ROSATP3 or ROSATWGA catalogue                    \\
 16  & var        &  variability flag                                         \\
     & ?hr2       &  {\bf or} error of hardness ratio 2
                        if entry is from ROSATP3 or ROSATWGA catalogue                    \\
 17  & red-sh.    &  red-shift                                                \\
     & ext, exl   &  {\bf or} source extent in arcsec and likelihood of source extent 
                        if entry is from ROSATP3 catalogue                    \\
 18  & dist       &  distance in kpc                                          \\
     & snr        &  {\bf or} signal-to-noise-ratio if FIRST entry \\
     & srcl, exp  &  {\bf or} likelihood (ROSATP3) or signal-to-noise-ratio 
                     (ROSATWGA/ROSHRI) of source 
                     detection and exposure in sec if entry is from ROSAT 
                     pointing catalogues \\
 19  & IR-12mu    &  infrared flux in the 12\mum\ band
                        if entry is from IRAS catalogues IRASFSC and IRASPSC  \\
     & R-f6cm     &  {\bf or} Radio flux in the 6cm band 
                        if entry is from VERONV8 catalogue                    \\
     & Ned-flux   &  {\bf or} Radio (IR) flux density in mJy
                        in the frequency (wavelength in microns) band 
                        given in column 20 if entry is from NED \\
 20  & IR-25mu    &  infrared flux in the 25\mum\ band
                        if entry is from IRAS catalogues IRASFSC and IRASPSC  \\
     & R-f11cm    &  {\bf or} Radio flux in the 11cm band 
                        if entry is from VERONV8 catalogue                    \\
     & fl-band    &  {\bf or} frequency (wavelength in microns) band used for
                      flux given in column 19 if entry is from NED \\
     & src-cps    &  {\bf or} source count-rate (vignetting corrected)
                        in the broad band in counts/sec
                        if entry is from ROSAT pointing catalogues                  \\
 21  & IR-60mu    &  infrared flux in the 60\mum\ band
                        if entry is from IRAS catalogues IRASFSC and IRASPSC  \\
     & R-f20cm(I) &  {\bf or} Radio integrated flux density 
                        in mJy in the 20cm band
                        if entry is from FIRST or NVSS catalogues             \\
     & ?src-cps   &  {\bf or} error of source count-rate in the broad band
                        in counts/sec 
                        if entry is from ROSAT pointing catalogues                 \\
 22  & IR-100mu   &  infrared flux in the 100\mum\ band
                        if entry is from IRAS catalogues IRASFSC and IRASPSC  \\
     & R-f20cm(P) &  {\bf or} Radio peak flux density in mJy
                        in the 20cm band 
                        if entry is from FIRST or NVSS catalogues             \\
     & size       &  {\bf or} size of object in arcmin if entry is from NED        \\
     & roff       &  {\bf or} off-axis angle in arcmin 
                        if entry is from ROSAT pointing catalogues                  \\
 23  & comments   &  additional information concerning type or class of object\\
 24  & orgdat     &  date (yymmdd) when the source was included \\
\strutb
 25  & moddat     &  date (yymmdd) when the source properties were changed \\
\hline
\end{tabular}
}\end{center}

\end{table*}

\def\aa{A\&A}
\def\aas{A\&AS}
\def\aca{Acta Astron.}
\def\apj{ApJ}
\def\apjs{ApJS}
\def\asj{AJ}
\def\ibvs{Inf. Bull. of Var. Stars}
\def\mnras{MNRAS}
\def\ea{{et al.}}
\def\asp{Astron. Soc. of the Pac.}
\def\pasp{PASP}
{}
\end{document}